\documentclass[aps,prl,reprint,superscriptaddress,longbibliography,footinbib,nobibnotes,floatfix]{revtex4-2}

\usepackage{graphicx}%
\usepackage{CJKutf8}
\usepackage{dcolumn}
\usepackage[nointegrals]{wasysym} %
\usepackage{bm}%

\usepackage[T1]{fontenc}
\usepackage[utf8]{inputenc}
\usepackage{lmodern}
\usepackage{amsmath,amssymb,bm,graphicx,xcolor,microtype}
\usepackage[colorlinks=true,allcolors=blue!55!black]{hyperref}
\graphicspath{{figures/}}
\newcommand{\bk}{\mathbf k}
\newcommand{\bq}{\mathbf q}
\newcommand{\br}{\mathbf r}
\newcommand{\bG}{\mathbf G}
\newcommand{\bC}{\mathbf C}
\newcommand{\bn}{\mathbf n}
\newcommand{\dd}{\mathrm d}
\newcommand{\ii}{\mathrm i}
\usepackage[normalem]{ulem}
\newcommand{\Tr}{\operatorname{Tr}}
\usepackage{booktabs,placeins}

\begin{document}

\title{Three-dimensional quantum Hall insulators in magnetic pyrochlore systems}
\author{Ze-Xun Lin(\begin{CJK}{UTF8}{gbsn}林泽勋\end{CJK})}
\email{zl659@cam.ac.uk}
\affiliation{TCM group, Department of Physics, University of Cambridge, Cambridge CB3 0US, United Kingdom}
\author{Robert-Jan Slager}
\affiliation{Department of Physics and Astronomy, The University of Manchester, Manchester M13 9PL, United Kingdom}
\author{Claudio Castelnovo}
\affiliation{TCM group, Department of Physics, University of Cambridge, Cambridge CB3 0US, United Kingdom}
\date{\today}

\begin{abstract}
We show that a class of pyrochlore materials provides a route to realize 3D Quantum Hall insulators, based on magnetic texture and flat band interference. We study itinerant electrons on a pyrochlore lattice Kondo-coupled to localized moments living on another interpenetrating pyrochlore lattice, with the moments forming a field-selected three-in/one-out texture. The noncoplanar texture generates complex spin-dependent hopping amplitudes and unequal local exchange energies, reconstructing the compact-localized parent flat bands into narrow bands separated by a global three-dimensional gap. The low-energy states are concentrated predominantly on kagome planes within the three-dimensional lattice, reflecting the combined effects of quantum interference and the texture-induced sublattice potential.
When the lowest band is fully occupied, the response is characterized by a Chern triplet $\bC=(1,1,1)$ and reciprocal Hall vector $\bG_{\mathrm H}=(2\pi/a)(1,1,1)$, giving $\sigma_{yz}=\sigma_{zx}=\sigma_{xy}=e^2/(ha)$. The quantized bulk Hall response enforces chiral surface spectral flow, visible as noncontractible Fermi loops at the surface. Our work opens a new direction in the search for three-dimensional quantum Hall physics in magnetic pyrochlore systems, and suggests a broad design principle based on the interplay of magnetic texture and flat-band interference.
\end{abstract}
\maketitle
 
Quantum Hall (QH) physics is usually regarded as a two-dimensional phenomenon. A strong magnetic field applied to a three-dimensional electron gas does not generally produce a QH insulator, since Landau levels remain dispersive along the field direction. A crystal can evade this problem by opening a gap in all three reciprocal directions, allowing a quantized Hall response indexed by a reciprocal-lattice vector~\cite{Halperin1987,KohmotoHalperinWu1992}. One established route is to start from QH layers and introduce interlayer coupling, yielding three-dimensional QH phases adiabatically connected to stacks of two-dimensional QH states~\cite{bernevig2007theory, kruthoff2017}. Such constructions, however, leave open the question whether all necessary ingredients can instead emerge within an intrinsically three-dimensional lattice.

\begin{figure*}[!tbp]
\centering
\includegraphics[width=0.985\textwidth]{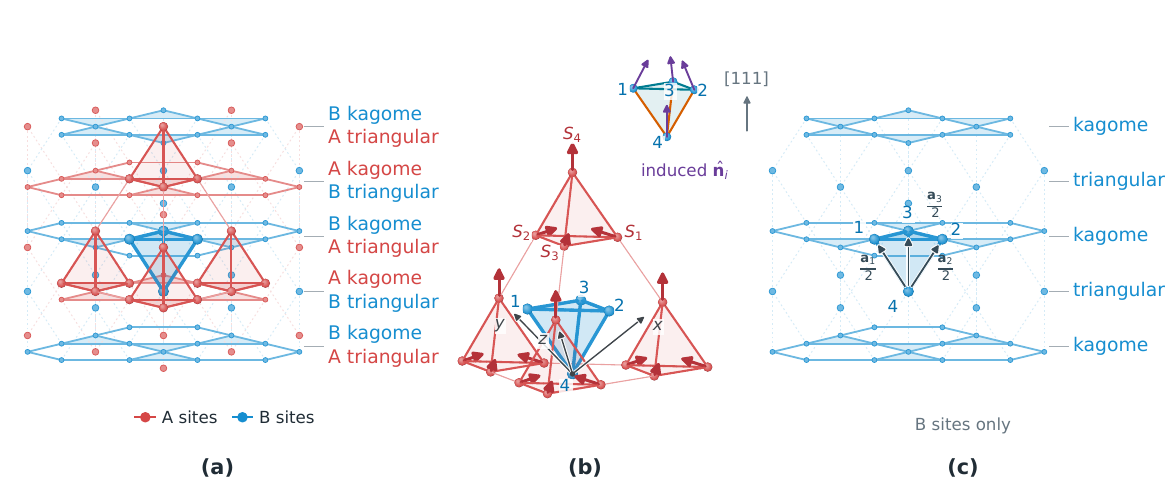}
\caption{Lattice structure of A$_2$B$_2$O$_7$. (a) Interpenetrating pyrochlore lattices of localized A-site moments (red) and itinerant B-site electrons (blue). The highlighted motif contains four A-tetrahedra surrounding a B-tetrahedron. (b) Three-in/one-out A-site moments (red arrows), global cubic axes, and the resulting B-site exchange-field directions (inset). The numbers indicate the four sublattices of each pyrochlore lattice. (c) Viewed along $[111]$, the B lattice alternates between kagome planes (sites 1--3) and triangular planes (site 4).}
\label{fig:geometry}
\end{figure*}

In this letter we discuss such a possibility in which magnetic texture generates a quantized Hall response 
where the (quasi-)layered QH physics is emergent rather than assumed from the outset. 
Our model and mechanism are inspired by pyrochlore materials A$_2$B$_2$O$_7$, where the A-sites host large localized $f$-electron moments and the B-sites host itinerant $d$-electrons on an interpenetrating pyrochlore lattice [Fig.~\ref{fig:geometry}]~\cite{Matsuhira2011}. In a nearest-neighbor tight-binding description, the B-site electrons exhibit flat bands consisting of compact localized states~\cite{Mielke1992,Bergman2008,WangJiangYao2024}. Kondo coupling to a noncoplanar A-site texture~\cite{Machida2009} modifies these states into weakly dispersive bands. We find that the texture opens a global gap, and filling the lowest band produces a three-dimensional QH insulator without Landau quantization.

Pyrochlore systems in general support itinerant noncoplanar order~\cite{Chern2010}, chirality-driven anomalous Hall effects~\cite{IshizukaMotome2013,UdagawaMoessner2013,IshizukaMotome2014}, quadratic band contacts~\cite{MoonXuKimBalents2013}, topological insulators~\cite{GuoFranz2009,YangKim2010}, Weyl semimetals~\cite{WanTurnerVishwanathSavrasov2011,YangLuRan2011,GoswamiRoyDasSarma2017}, and thickness-enhanced Hall responses~\cite{YangNagaosa2014}. 
External fields can be used to select or reorganize large-moment $f$-electron magnetic textures, which in turn affect itinerant $d$-electrons in ways that cannot be attained by the external fields themselves~\cite{Sakakibara2003,Ueda2015, Lefrancois2017, Cathelin2020, Pearce2022,  Coak2024}. 
Here the combination of this magnetic control and flat-band interference gives a fully gapped bulk Hall phase, together with a chiral surface metal. 

\emph{Magnetic texture and tight-binding model.---}
We consider one spinful itinerant orbital per B-site,
$c_i=(c_{i\uparrow},c_{i\downarrow})^T$, coupled to quasi-static A-site moments of fixed length $S$. We work in an isotropic approximation in which each B site couples equally to its six nearest-neighbor A-site moments. For the nonlocal exchange, we likewise adopt a common amplitude for all B-site pairs connected through an intermediate nearest-neighbor A site.
\begin{align}
\hat H={}&-\sum_{\langle ij\rangle}
\left(c_i^\dagger T_{ij}c_j+\mathrm{H.c.}\right)
\nonumber\\
&-J_K\sum_i\sum_{\ell\in\mathcal N_A(i)}
c_i^\dagger(\mathbf S_\ell^A\cdot\bm\sigma)c_i
\nonumber\\
&-J_K'\sum_{\ell\in A}
\sum_{\substack{i,j\in\mathcal N_B(\ell)\\i<j}}
\left[
c_i^\dagger(\mathbf S_\ell^A\cdot\boldsymbol\sigma)c_j
+\mathrm{H.c.}\right]
\label{eq:KondoH}
\end{align}
Here $\mathcal N_A(i)$ denotes the six nearest-neighbor A-sites forming a planar hexagon centered on B-site $i$, and $\mathcal N_B(\ell)$ denotes the six nearest-neighbor B-sites around A-site $\ell$.
The couplings $J_K$ and $J_K'$ describe onsite Kondo exchange and nonlocal Kondo exchange~\cite{Kirikoshi2026}, respectively. The hopping matrix is $T_{ij}=t_{ij}I_2+\ii\lambda_{ij}\boldsymbol\nu_{ij}\cdot\bm\sigma$, where $\boldsymbol\nu_{ij}$ specifies the bond-dependent spin--orbit coupling (SOC) axis. The strength and spatial structure of the nonlocal exchange depend on the microscopic orbital hybridization. We focus on the onsite exchange ($J_K'=0$) and neglect SOC ($\lambda_{ij}=0$) in the main text. In the Supplemental Material (SM)~\cite{SupplementalMaterial}, we show that the 
physics we uncover
is robust against SOC and weak nonlocal exchange. In particular with the latter we find that stronger coupling can close the indirect gap without necessarily closing the direct band gap.

For A-site moments with local $\langle111\rangle$ easy axes, as in
Ho or Dy pyrochlores, we consider the global $[111]$-field-selected
three-in/one-out texture in Fig.~\ref{fig:geometry}(b)~\cite{Pearce2022}.
This texture preserves primitive-cell translations.
For $J_K>0$, define the local exchange magnitude and direction by
$J_{{\rm ex},i}\bn_i
=J_K\sum_{\ell\in\mathcal N_A(i)}\mathbf S_\ell^A$,
with $J_{{\rm ex},i}>0$ and $|\bn_i|=1$.
The effective local Zeeman potentials $\bn_i$ can only take one of four sublattice-dependent values~\cite{SupplementalMaterial}: 
\begin{equation}
    \bn_i \in \left\{ 
    \begin{array}{ll}
    (3,1,1)/\sqrt{11} & {\rm sublattice \; 1} \\ 
    (1,3,1)/\sqrt{11} & {\rm sublattice \; 2} \\ 
    (1,1,3)/\sqrt{11} & {\rm sublattice \; 3} \\ 
    (1,1,1)/\sqrt{3} & {\rm sublattice \; 4} 
    \end{array}
    \right. 
    \, , 
	\label{eq:textureassignment}
\end{equation}
see Fig.~\ref{fig:geometry}(b). 
Although $J_K$ is uniform, the vector sums give unequal exchange
magnitudes:
$J_{{\rm ex},1,2,3}=J=2J_KS\sqrt{11/3}$ and
$J_{{\rm ex},4}=rJ$, with $r=\sqrt{3/11}$.
Sites 1--3 form the kagome planes and site 4 the triangular planes
in the $[111]$ decomposition of the pyrochlore lattice illustrated in Fig.~\ref{fig:geometry}.

While the topological properties uncovered in our work are borne by the full spin model in Eq.~\eqref{eq:KondoH}, they are most transparently understood in a strong-coupling projected limit that we present hereafter. Calculations for the original Hamiltonian are presented in the Supplemental Material~\cite{SupplementalMaterial}. They establish the persistence of the insulating phase beyond the projected limit and quantify the finite-exchange corrections to the bulk gap.

In the strong-exchange limit
$J_{{\rm ex},i}\gg |t_{ij}|,|\lambda_{ij}|$,
we project onto the locally aligned states
$(\bn_i\cdot\bm\sigma)|\chi_i^+\rangle=|\chi_i^+\rangle$.
To leading order, this gives one spinless orbital per B-site, which results in a particularly simple model:
\begin{align}
\hat H_4={}&-\sum_{\langle ij\rangle}
\left(\zeta_{ij}f_i^\dagger f_j+\mathrm{H.c.}\right)
-
\sum_iJ_{{\rm ex},i}f_i^\dagger f_i,
\label{eq:mainprojectedhopping}
\end{align}
where $\zeta_{ij}=\langle\chi_i^+|T_{ij}|\chi_j^+\rangle$.
Without SOC, $\zeta_{ij}=t_{ij}u_{ij}$, with
$u_{ij}=\langle\chi_i^+|\chi_j^+\rangle$ and
$|u_{ij}|=\cos(\theta_{ij}/2)$, where $\theta_{ij}$ is the angle
between the local exchange directions.
As an electron hops around a triangular face of a tetrahedron, the Berry phase it gains is half the solid angle subtended by the three directions $\bn_i$ on the Bloch sphere. This makes transparent how noncoplanar moments induce electronic Berry phases, generalizing the chiral kagome mechanism~\cite{Ohgushi2000} to three dimensions. 

\emph{From compact localized states to a 3D Chern band.---}
For a uniform collinear texture, the projected nearest-neighbor pyrochlore model has two degenerate flat bands associated with compact localized states~\cite{Mielke1992}. In these states, amplitudes on a closed loop interfere destructively at neighboring sites, preventing electrons from propagating outside the loop. We choose equal negative bare hoppings, $t_{ij}=t < 0$, placing the flat bands at the bottom of the spectrum. On the non-bipartite pyrochlore lattice, this sign choice cannot be removed by a site-dependent gauge transformation and is a physical model assumption.

\begin{figure}[!tbp]
\centering
\includegraphics[width=0.9\columnwidth]{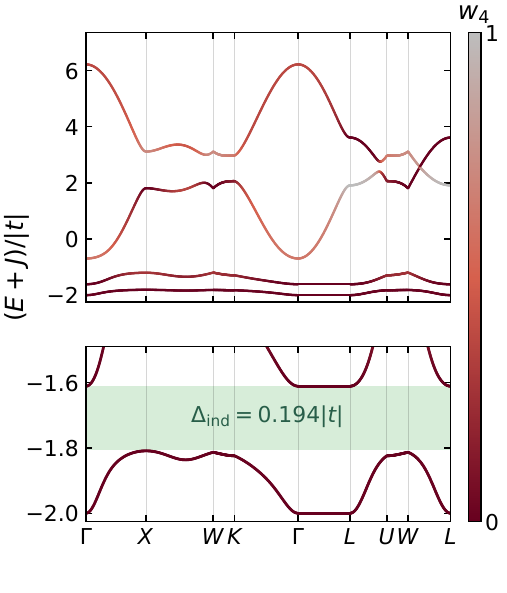}
\caption{Projected bands for $J=4|t|$, $J_{{\rm ex},4}=rJ$, $t_{ij}=t < 0$, and $\lambda_{ij}=0$. The lower panel enlarges the lowest-band gap. Colors give the site-$4$ probability. Energies are shifted by $+J$ for convenience. These parameters are also used in Figs.~\ref{fig:surface} and~\ref{fig:fermiriver}.}
\label{fig:modelbands}
\end{figure}

\begin{figure*}[!tbp]
\centering
\includegraphics[width=0.985\textwidth]{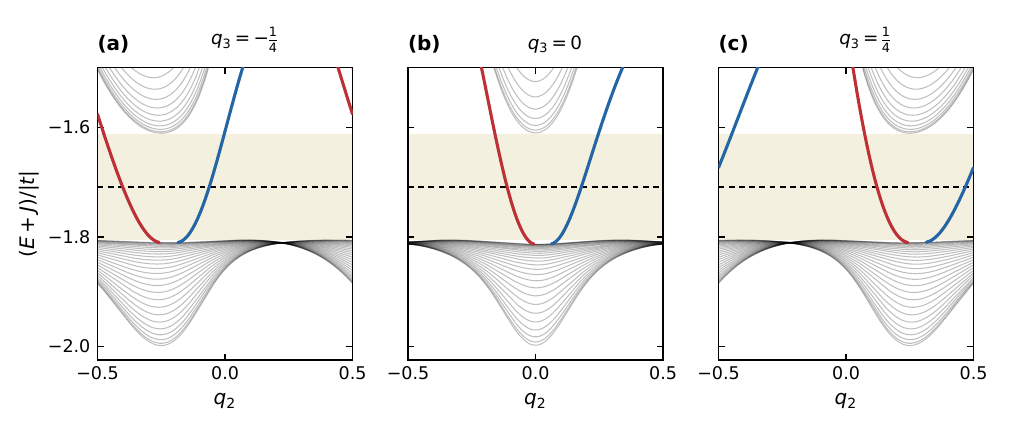}
\caption{Finite-slab energy spectrum at $q_3=-1/4,0,1/4$ for a slab composed of $N=28$ unit-cell layers parallel to the $234$ plane. Gray curves are the slab bands. Red and blue highlight edge states, identified here as band states that have at least half their probability in the outermost three unit-cell layers 
(top: red; bottom: blue; with respect to the axis perpendicular to the $234$ plane). Pale-yellow shading marks the bulk indirect gap; the dashed line is its midpoint, $\widetilde\mu\simeq-1.708|t|$.}
\label{fig:surface}
\end{figure*}

Upon introducing the magnetic texture induced by the applied field in the global $[111]$ direction, the hopping acquires the spin-overlap factors in Eq.~\eqref{eq:mainprojectedhopping}. Although the bare hopping remains equal on every bond, the projected hopping magnitudes and phases depend on the local exchange directions. This changes the interference conditions responsible for the collinear flat bands and gives rise to weakly dispersive low-energy bands.

At the same time, the aligned state on site 4 gains a smaller exchange energy, $rJ$, than the states on sites 1--3. After a common energy shift $\epsilon_n=E_n+J$ and $\widetilde\mu=\mu+J$, the kagome-site energies are zero and the triangular-site energy is $\delta=(1-r)J$. This relative onsite energy is fixed by the magnetic texture, biasing toward less occupation on the triangular sites. The complex hoppings and unequal onsite energies together lift the degeneracy and open a global gap. 

Figure~\ref{fig:modelbands} shows the projected spectrum at $J=4|t|$, for which $\delta\simeq1.911|t|$, with an indirect gap of $\Delta_1^{\rm ind}=\min_{\bk}E_{2\bk}-\max_{\bk}E_{1\bk}\simeq0.1944|t|$. The lowest-band width is $W_1=\max_{\mathbf k}E_{1\mathbf k}-\min_{\mathbf k}E_{1\mathbf k}\simeq0.1944|t|$, giving $\Delta^{\mathrm{ind}}_1/W_1\simeq1$. For the same parameters, the original Hamiltonian retains an indirect gap of approximately $0.0581|t|$~\cite{SupplementalMaterial}. One electron per primitive cell fills an isolated lowest band. Its Hall response follows from the Berry-curvature form of the Kubo formula~\cite{Halperin1987, KohmotoHalperinWu1992, SupplementalMaterial},
\begin{equation}
    \sigma_{\alpha\beta}=-\frac{e^2}{\hbar}
    \sum_{n\,\mathrm{occ.}}\int_{\rm BZ}
    \frac{\dd^3k}{(2\pi)^3}
    \Omega^n_{\alpha\beta}(\bk)
    \, ,
    \label{eq:kubo}
\end{equation}
with Berry curvature $\Omega^n_{\alpha\beta}=-2\,\mathrm{Im}\langle\partial_{k_\alpha}u_{n\mathbf k}| \partial_{k_\beta}u_{n\mathbf k}\rangle$, spatial indices in Cartesian coordinates $\alpha,\beta,\gamma \in \{x,y,z\}$, and $|u_{n\mathbf k}\rangle$ being the cell-periodic part of the Bloch state. Writing $\bk=\sum_aq_a\mathbf b_a$ in primitive reciprocal vectors $\mathbf b_a$, define $\Omega^{(\boldsymbol{q})}_{bc}=b_{b,\alpha}b_{c,\beta}\Omega^1_{\alpha\beta}$ for the occupied band. The Chern numbers for the lowest band are given as
\begin{equation}
C_a=-\frac{1}{2\pi}\int_0^1\!\dd q_b\int_0^1\!\dd q_c\,\Omega^{(\boldsymbol{q})}_{bc},
\label{eq:chern}
\end{equation}
with $(a,b,c)$ cyclic. At zero temperature, with the chemical potential in the gap, the Hall response is quantized in units of the
primitive reciprocal-lattice vectors~\cite{Halperin1987, KohmotoHalperinWu1992}, 
\begin{equation}
\sigma^{\rm H}_{\alpha\beta}=\frac{e^2}{2\pi h}\epsilon_{\alpha\beta\gamma}G_{{\rm H},\gamma},
\qquad \bG_{\rm H}=\sum_{a=1}^3C_a\mathbf b_a,
\label{eq:quantized}
\end{equation}
For the parameters considered, $\mathbf{C}=(1,1,1)$ and
$\mathbf{G}_{H}=(2\pi/a)(1,1,1)$, where $a$ is the cubic
lattice constant, giving
$\sigma_{yz}=\sigma_{zx}=\sigma_{xy}=e^2/(ha)$.
Wilson-loop spectral flow independently confirms the
three integers~\cite{SupplementalMaterial}.

\begin{figure}[!tbp]
\centering
 \includegraphics[width=\columnwidth]{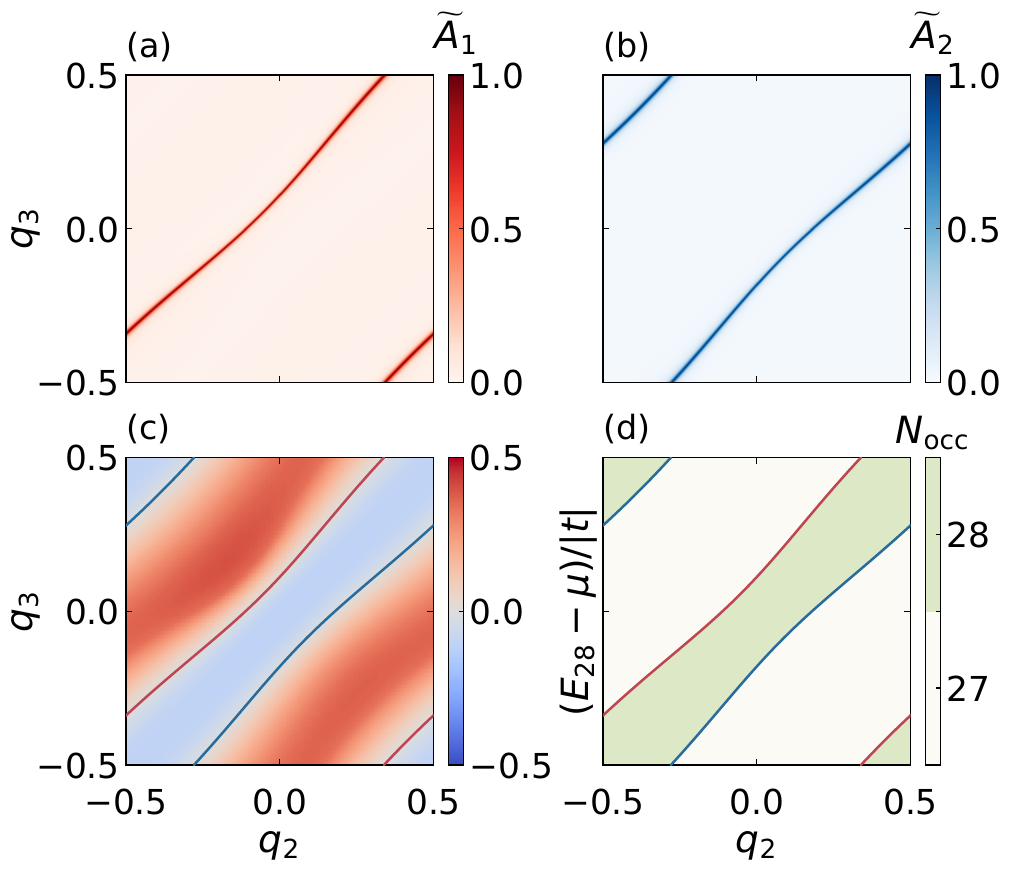}
\caption{Surface Fermi loops and Fermi river at the gap midpoint. (a,b) Normalized near-$\mu$ spectral intensity on the two faces. (c) Signed energy $E_{28}-\mu$, with the two face-resolved zero contours. (d) Occupied-level count: the $N_{\rm occ}=28$ strip is bounded by these loops, while the complement has $N_{\rm occ}=27$. Opposite edges are identified; each loop winds geometrically as $(1,1)$ on the surface Brillouin-zone torus.}
\label{fig:fermiriver}
\end{figure}

Topologically, this state is in the same class as stacked QH layers. Microscopically, it is realized on a genuinely three-dimensional lattice with equal bare hopping on every bond. Interestingly, the spatial distribution of the lowest quasi-flat band in Fig.~\ref{fig:modelbands} reveals an emergent quasi-layered character\footnote{Interference-induced quasi-layered wave functions have previously been identified in different pyrochlore settings
~\cite{Trescher2012,Bergholtz2015}.}: the electron weight is concentrated predominantly within the kagome planes, while the triangular-site weight is nonzero but small. This behavior appears to be underpinned by three central ingredients: the compact-localized flat bands in the uniform nearest-neighbor tight-binding problem; the interference effects due to the Berry phases introduced by the coupling of the conduction electrons to the background magnetic texture; and the fact that the latter also induces a different chemical potential on one sublattice ($4$) with respect to the others ($1,2,3$). In the SM~\cite{SupplementalMaterial} we investigate the effects of these ingredients separately. We find that interference is primarily responsible for the quasi-layered nature of the states. Moreover, without this interference, the gap remains closed even in the limit of vanishing interlayer hopping $t'$~\cite{SupplementalMaterial}. While the effective sublattice potential also induces a layered periodicity in the electronic density, its effect is far less substantial; on the other hand, the sublattice potential is in turn key to open a full 3D gap between the lowest and first excited bands (interference alone leads to quadratic band touching at the $\Gamma$ point)~\cite{SupplementalMaterial}. 
Our results provide an intuitive connection between the conventional stacked-layer picture of 3D QH and its realization here in a three-dimensional magnetic crystal.

\emph{Quantized bulk response and chiral surface metal.---}
Since the bulk is gapped, its long-wavelength Hall response is governed by a $(3+1)$D crystalline Chern--Simons action~\cite{NissinenVolovik2019}. The corresponding current is $\mathbf j_{\rm H}=(e^2/2\pi h)\mathbf E\times\bG_{\rm H}$. The action and its equivalent local form with a spatially winding axion angle are given in the SM. For a surface with outward normal $\hat{\mathbf m}$, charge conservation requires
\begin{align}
\partial_t\rho_{\rm surf}+\boldsymbol\nabla_\parallel\cdot\mathbf j_{\rm surf}
&=\hat{\mathbf m}\cdot\mathbf j_{\rm H}\nonumber\\
&=\frac{e^2}{2\pi h}\mathbf E\cdot(\bG_{\rm H}\times\hat{\mathbf m}).
\label{eq:surfacecontinuity}
\end{align}
For \(\mathbf G_H\times\hat{\mathbf m}\neq0\), an electric field along $\mathbf G_H\times\hat{\mathbf m}$ produces a nonzero Hall-current inflow at the boundary. The corresponding spectral flow requires gapless surface states in the noninteracting system. 
Thus surfaces whose normals are not parallel to
$\mathbf{G}_{H}$ support a chiral surface
metal~\cite{BalentsFisher1996}. In the presence of disorder, the conventional chiral-metal
regime can exhibit ballistic propagation along the chiral
direction and diffusive motion in the transverse surface
direction~\cite{ChoBalentsFisher1997}.

To illustrate this bulk--surface correspondence, we calculate
the spectrum of a slab parallel to the 234 kagome face.
The slab is periodic along $\mathbf{a}_2$ and $\mathbf{a}_3$,
with conserved surface momenta $(q_2,q_3)$, and contains
$N=28$ primitive-cell layers along $\mathbf{a}_1$,
corresponding to 28 kagome and 28 triangular layers.

Fig.~\ref{fig:surface} gives some example views of the finite-slab eigenenergies in the
projected model. Each panel shows a fixed $q_3$ cut. Within it, the chemical potential crosses one in-gap
state on each face, and the two states propagate oppositely, as required
by the opposite outward normals.

All the $q_3$ cuts are combined in a 2D plot in Fig.~\ref{fig:fermiriver}, which gives the global information over the complete $(q_2,q_3)$ surface Brillouin zone at $\widetilde\mu=-1.708|t|$. This behavior leads to immediate spectroscopic evidence: at either surface there is a noncontractible Fermi loop in principle measurable by ARPES. Together the loops form the boundary of a strip within which one additional slab level is occupied, forming a Fermi river. If the chemical potential lies above the topological gap, surface-state contours within that gap may remain accessible to ARPES at finite binding energy, even when the filling does not support quantized Hall transport. Such contours should be distinguished from Fermi loops at the actual chemical potential. 
Materials that stabilize the required magnetic texture without an external field would be particularly promising for ARPES measurements. The surface-weight and spectral-intensity definitions are given in the SM~\cite{SupplementalMaterial}. 

\emph{Robustness and materials realization.---}
As already mentioned earlier, our central results are best illustrated in the strong-exchange limit. However, the results remain qualitatively unchanged if one considers the original Hamiltonian instead, as demonstrated in the SM~\cite{SupplementalMaterial}. 
There we discuss the SOC stability in detail and present the phase diagram as a function of $J$, identifying the range of $J/|t|$ over which the full
spinful model retains a positive indirect gap and
$\mathbf{C}=(1,1,1)$. Hartree--Fock calculations with repulsive Coulomb interactions
also retain the $\mathbf{C}=(1,1,1)$ insulating state at full
filling of the lowest band over the interaction range studied
(see SM~\cite{SupplementalMaterial}).

Rare-earth iridate pyrochlores provide a natural material platform for this mechanism. In Ho$_2$Ir$_2$O$_7$ or Dy$_2$Ir$_2$O$_7$, for instance, large anisotropic rare-earth moments are known to generate an exchange field for the Ir electrons. In turn, a $[111]$ field can select the relevant moment texture~\cite{Pearce2022}. Our model can be viewed as a minimal single-orbital description of the Ir electrons in these materials, coupled to the magnetic texture (but neglecting the direct orbital and Zeeman couplings of the electrons to the applied field).

The 3D QH phase can be probed through its quantized bulk Hall tensor and chiral surface metallic behavior. ARPES can be used to detect the noncontractible surface Fermi loops, while surface-sensitive transport may provide complementary evidence~\cite{DruistTurley1998}. 

\emph{Discussion and outlook.---}
We have demonstrated a three-dimensional QH insulator arising from the interplay of non-coplanar magnetic textures and pyrochlore band interference. %
Our results reveal two complementary roles of the magnetic texture. The induced complex hopping amplitudes reconstruct the compact-localized states through quantum interference, producing the emergent quasi-layered character of the lowest band. The accompanying sublattice-dependent exchange energies lift the remaining quadratic band touching at the $\Gamma$ point and establish a global insulating gap. Neither effect requires an explicitly layered bare hopping structure. Thus, while the resulting QH phase is topologically equivalent to stacked two-dimensional QH layers, its microscopic realization emerges from the interference and magnetic texture of a fully three-dimensional lattice. 

When filling the lowest Chern band, the system realizes a bulk quantized response and a chiral surface metal. At partial filling, the relatively flat Chern bands inherited from compact localized states provide an interesting setting for exploring strongly correlated phases. Whether interactions can stabilize intrinsically three-dimensional fractional QH states~\cite{Naud2000, LevinFisher2009} in this microscopic model remains an open question. 
Beyond the static-moment approximation, moment dynamics will introduce fluctuations of both the effective hopping and local exchange energies, whose effects are left for future investigation. 
More broadly, our results suggest magnetic texture and lattice interference as ingredients for realizing three-dimensional quantum Hall and topological phenomena in bulk materials. 

\emph{Acknowledgments.---}
We are grateful to Emil J. Bergholtz for insightful discussions. GPT-5.6 Sol and GPT-6 Astra (OpenAI) were used to assist with code development and debugging; Z.-X.L. reviewed and revised the AI-assisted code.
This work was supported in part by the Engineering and Physical Sciences Research Council (EPSRC) grant No.~EP/V062654/1 (Z-X.L. and C.C.). R.-J.S.~acknowledges funding from an EPSRC ERC underwrite grant EP/X025829/1.

\bibliographystyle{apsrev4-2}
\bibliography{references}

\clearpage
\onecolumngrid
\graphicspath{{SI_figures/}}
\setcounter{section}{0}
\setcounter{subsection}{0}
\setcounter{equation}{0}
\setcounter{figure}{0}
\setcounter{table}{0}
\renewcommand{\theHsection}{supp.\arabic{section}}
\renewcommand{\theHsubsection}{\theHsection.\arabic{subsection}}
\renewcommand{\theHequation}{supp.\arabic{equation}}
\renewcommand{\theHfigure}{supp.\arabic{figure}}
\renewcommand{\theHtable}{supp.\arabic{table}}
\renewcommand{\thesection}{S\Roman{section}}
\setcounter{secnumdepth}{3}
\renewcommand{\theequation}{S\arabic{equation}}
\renewcommand{\thefigure}{S\arabic{figure}}
\renewcommand{\thetable}{S\arabic{table}}

\begin{center}
{\large\bfseries Supplemental Material for "Three-dimensional quantum Hall insulators in magnetic pyrochlore systems"\par}
\medskip
Ze-Xun Lin$^{1}$, Robert-Jan Slager$^{2}$, Claudio Castelnovo$^{1}$\par
\smallskip
{\small $^{1}$TCM group, Department of Physics, University of Cambridge, Cambridge CB3 0US, United Kingdom\par}
{\small $^{2}$Department of Physics and Astronomy, The University of Manchester, Manchester M13 9PL, United Kingdom\par}
\smallskip
{\small (Dated: \today)\par}
\end{center}

\section{Geometry, magnetic texture, and unequal exchange}
\label{sec:Smodel}
In this section we provide further details on the lattice geometry and the magnetic texture shown in Fig.~1
of the main text, and we derive explicitly the exchange field directions and magnitudes exerted on a B-site electron by its nearest-neighboring A-site moments ($\vert S_\ell^A \vert =S$), used in the main-text Eqs.~(1) and~(2). 

The face-centered-cubic primitive translations and basis of the itinerant B-sites pyrochlore lattice are
\begin{align}
\mathbf a_1&=\tfrac a2(0,1,1),&\mathbf a_2&=\tfrac a2(1,0,1),&\mathbf a_3&=\tfrac a2(1,1,0),\nonumber\\
\mathbf b_1&=\tfrac{2\pi}{a}(-1,1,1),&\mathbf b_2&=\tfrac{2\pi}{a}(1,-1,1),&\mathbf b_3&=\tfrac{2\pi}{a}(1,1,-1),\nonumber\\
\mathbf r_1&=\mathbf a_1/2,&\mathbf r_2&=\mathbf a_2/2,&\mathbf r_3&=\mathbf a_3/2,\qquad\mathbf r_4=0.
\label{eq:Sgeometry}
\end{align}
Here $a$ is the cubic lattice constant, and the nearest-neighbor pyrochlore separation is $a/\sqrt8$; $\mathbf a_i\cdot\mathbf b_j=2\pi\delta_{ij}$, and $\mathbf a_i$ and $\mathbf b_i$ denote the primitive and reciprocal lattice vectors; the primitive volume is $a^3/4$; and the crystal momentum is parameterized as $\bk=\sum_iq_i\mathbf b_i$. Sites $1,2,3$ form the kagome set in the $[111]$ decomposition; sites $4$ form interleaving triangular lattices (see Fig.~1 in the main text). The Cartesian axes are along $[100]$, $[010]$, and $[001]$; $[111]$ is the cubic body diagonal. 

\begin{figure}[!tbp]
\centering
\includegraphics[width=0.66\textwidth]{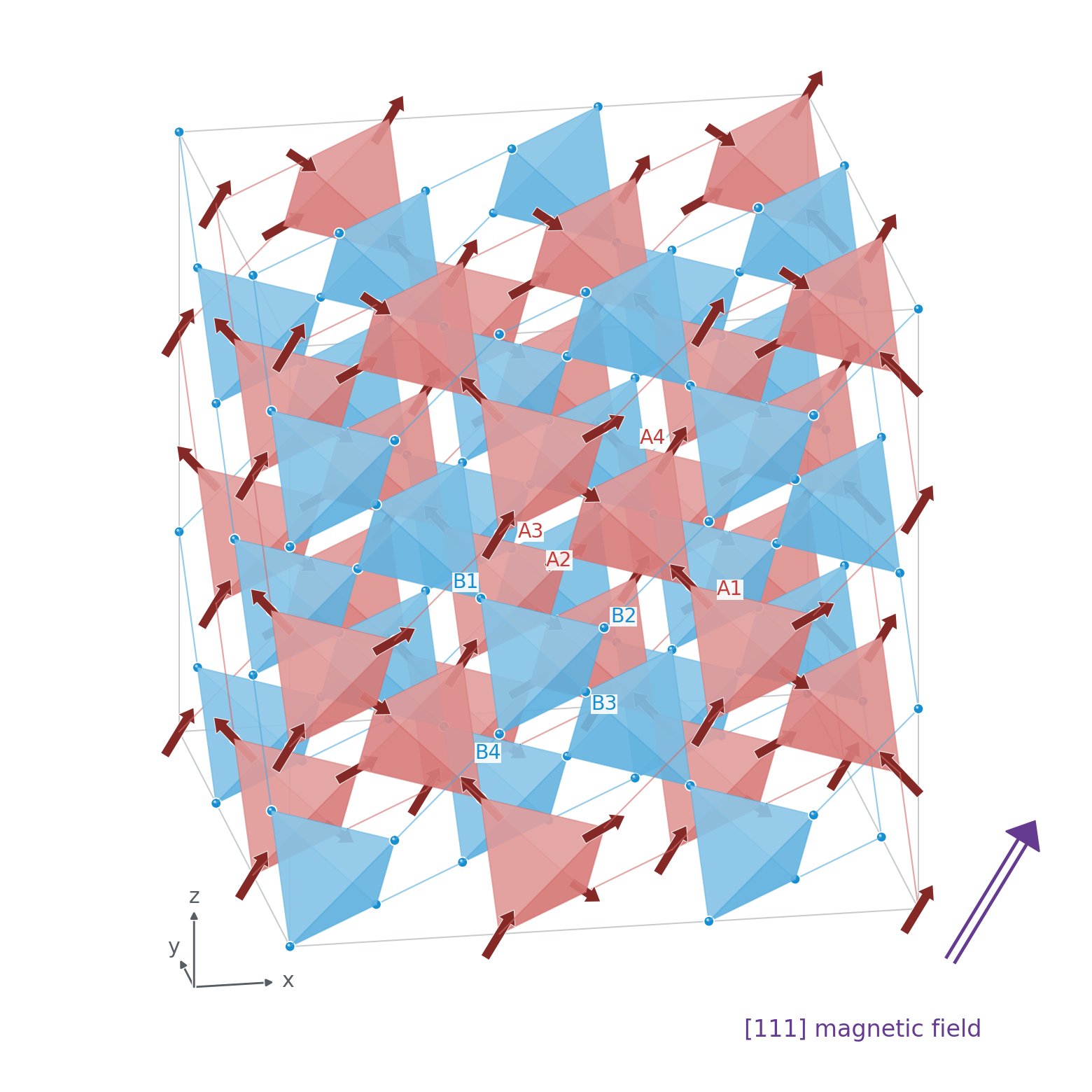}
\caption{Interpenetrating A-site (red) and B-site (blue) pyrochlore lattices. Dark red arrows show the local Ising moments of Eq.~\eqref{eq:SAmoments}; the purple double-line arrow indicates the global $[111]$ direction of the system magnetization. Labels A1--A4 and B1--B4 denote sublattice sites (called, without ambiguity,simply $1,2,3,4$ in the text), with $\mathbf S_4^A$ parallel to the global $[111]$. Summing the six neighboring A-site moments at each B-site gives Eq.~\eqref{eq:Sfields} and the normalized exchange directions $(\bm n_1,\bm n_2,\bm n_3,\bm n_4)$ (for the B-sites $1,2,3,4$, respectively).}
\label{fig:SABgeometry}
\end{figure}

The magnetic texture considered in the main text is shown in Fig.~\ref{fig:SABgeometry}. The two tetrahedral sublattices carry three-in/one-out and one-in/three-out configurations.
The A-site numbering follows again Fig.~1 of the main text: site $4$ is the triangular site on both networks, and sites $1,2,3$ are the kagome sites. For the displayed A-site tetrahedron, the outward local Ising axes are
\begin{equation}
\left\{ \hat{\mathbf e}_1,\hat{\mathbf e}_2,\hat{\mathbf e}_3,\hat{\mathbf e}_4 \right\}
=\frac1{\sqrt3}\left\{\vphantom{\sum}(1,-1,-1),(-1,1,-1),(-1,-1,1),(1,1,1) \right\}.
\end{equation}
The selected moments are $\mathbf S_i^A=-S\hat{\mathbf e}_i$ for $i=1,2,3$ and $\mathbf S_4^A=S\hat{\mathbf e}_4$, namely
\begin{equation}
(\mathbf S^A_1,\mathbf S^A_2,\mathbf S^A_3,\mathbf S^A_4)
=\frac S{\sqrt3}\bigl((-1,1,1),(1,-1,1),(1,1,-1),(1,1,1)\bigr).
\label{eq:SAmoments}
\end{equation}
Thus, $\mathbf S_4^A\parallel[111]$ is the outward moment. Each B site $i$ couples to six nearest A moments, namely two A moments from each of the other three sublattices, which gives
\begin{align}
\mathbf h^{A\to B}_i&=2\sum_{\ell\ne i}\mathbf S^A_\ell
,\nonumber\\
\left\{ \mathbf h^{A\to B}_1,\mathbf h^{A\to B}_2,\mathbf h^{A\to B}_3,\mathbf h^{A\to B}_4 \right\}
&=\frac{2S}{\sqrt3}\left\{\vphantom{\sum}(3,1,1),(1,3,1),(1,1,3),(1,1,1)\right\}.
\label{eq:Sfields}
\end{align}
For $J_K>0$, define $J_{{\rm ex},i}=J_K|\mathbf h_i^{A\to B}|$, $J=J_{{\rm ex},1}=J_{{\rm ex},2}=J_{{\rm ex},3}$ and $r=\sqrt{3/11}$. Then
\begin{equation}
J=2J_KS\sqrt{11/3},\qquad J_{{\rm ex},4}=2J_KS=rJ. 
\label{eq:Sprofile}
\end{equation}
Let us also introduce the normalized vectors 
$\{\bn_1,\allowbreak \bn_2,\allowbreak \bn_3,\allowbreak \bn_4\} = \{\mathbf h^{A\to B}_1/|\mathbf h^{A\to B}_1|,\allowbreak \mathbf h^{A\to B}_2/|\mathbf h^{A\to B}_2|,\allowbreak \mathbf h^{A\to B}_3/|\mathbf h^{A\to B}_3|,\allowbreak \mathbf h^{A\to B}_4/|\mathbf h^{A\to B}_4|\}$, where $\bm n_1=(3,1,1)/\sqrt{11}$, $\bm n_2=(1,3,1)/\sqrt{11}$, $\bm n_3=(1,1,3)/\sqrt{11}$, and $\bm n_4=(1,1,1)/\sqrt3$. For $J_K < 0$, the lower local spin is anti-aligned with these $\bn_i$. 
Therefore, for a fixed magnetic texture and $|J_K|$, reversing $J_K$ leaves the bulk spectrum unchanged but reverses the Chern vector and net surface chirality discussed in the main text. The numerical results here use $J_K>0$. 

\section{Hamiltonian and local-spin projection}
\label{sec:Sprojection}
We denote the eight-band Bloch matrix of the original
Hamiltonian $\hat H$ in Eq.~(1) of the main text as $H(\bk)$, which has four sublattices with spin degree of freedom. Its local spin projection gives the four-band
Bloch matrix $H_{\mathrm{proj}}(\bk)$ of the projected Hamiltonian
$\hat H_{\mathrm{proj}}$ in main-text Eq.~(3). 
We introduce a scalar onsite potentials $V_\alpha$ that will be useful for some of the investigations in the SM. The main text results have $V_\alpha=0$. 

\subsection{The original Bloch matrix}
In the Bloch basis $(1\uparrow,1\downarrow,\ldots,4\uparrow,4\downarrow)$, let $\alpha,\beta=1,\ldots,4$ label sublattices and $s,s'=\uparrow,\downarrow$ label spin. Quantum operators carry hats. For $J_K'=0$, the original Hamiltonian has matrix elements
\begin{align}
[H(\bk)]_{\alpha s,\alpha s'}&=V_\alpha\delta_{ss'}-J_{{\rm ex},\alpha}(\bn_\alpha\cdot\bm\sigma)_{ss'},\nonumber\\
[H(\bk)]_{\alpha s,\beta s'}&=-2(T_{\alpha\beta})_{ss'}\cos(\bk\cdot\mathbf d_{\alpha\beta}),\quad \alpha<\beta,\nonumber\\
[H(\bk)]_{\beta s',\alpha s}&=[H(\bk)]_{\alpha s,\beta s'}^*,\nonumber\\
T_{\alpha\beta}&=t_{\alpha\beta}I_2+\ii\lambda_{\alpha\beta}\boldsymbol\nu_{\alpha\beta}\cdot\bm\sigma,
\qquad \mathbf d_{\alpha\beta}=\mathbf r_\beta-\mathbf r_\alpha.
\label{eq:SH8}
\end{align}
Here $T_{\alpha\beta}$ is a $2\times2$ spin matrix, with $T_{\beta\alpha}=T_{\alpha\beta}^\dagger$ and $\boldsymbol\nu_{\beta\alpha}=-\boldsymbol\nu_{\alpha\beta}$. As in the main text, all bare hoppings are chosen to be uniform $t_{\alpha\beta}=t<0$, except where otherwise specified. 
For a collinear texture with equal exchange magnitudes, the choice of sign puts the flat bands at the bottom.

With $\mathbf c=\frac14\sum_{\alpha=1}^4\mathbf r_\alpha$, we choose the spin--orbit coupling (SOC) axes
\begin{equation}
\boldsymbol\nu_{\alpha\beta}=\frac{(\mathbf r_\alpha-\mathbf c)\times(\mathbf r_\beta-\mathbf c)}{|(\mathbf r_\alpha-\mathbf c)\times(\mathbf r_\beta-\mathbf c)|}.
\end{equation}
Explicitly, for $\alpha<\beta$,
\begin{align}
\boldsymbol\nu_{12}&=(1,1,0)/\sqrt2,&\boldsymbol\nu_{13}&=(-1,0,-1)/\sqrt2,&\boldsymbol\nu_{23}&=(0,1,1)/\sqrt2,\nonumber\\
\boldsymbol\nu_{14}&=(0,-1,1)/\sqrt2,&\boldsymbol\nu_{24}&=(1,0,-1)/\sqrt2,&\boldsymbol\nu_{34}&=(-1,1,0)/\sqrt2.
\end{align}
Sec.~\ref{sec:Ssoc} uses $\lambda_{\alpha\beta}=\lambda$ on every bond. This amplitude and the bond axes define the phenomenological SOC model.

\subsection{Local-spin projection}
We give the local-spin basis underlying the strong-exchange projection
in main-text Eq.~(3), valid for $\min_iJ_{{\rm ex},i}\gg |t|,|\lambda|$.
Here $i,j$ label real-space sites; their local fields and bond matrices have the sublattice values defined above.
Choose normalized eigenspinors $(\bn_i\cdot\bm\sigma)|\chi_i^\pm\rangle=\pm|\chi_i^\pm\rangle$, with
\begin{equation}
|\chi_i^+\rangle=(\cos(\vartheta_i/2),\ e^{\ii\varphi_i}\sin(\vartheta_i/2))^T,
\end{equation}
where $(\vartheta_i,\varphi_i)$ are the polar and azimuthal angles of $\bn_i$, and $(1,0)^T$ and $(0,1)^T$ denote spin-up and spin-down in the global $z$ basis. Writing $\hat c_i=|\chi_i^+\rangle \hat f_i+|\chi_i^-\rangle \hat g_i$, we project onto the locally aligned spin states to obtain
\begin{equation}
\hat H_{\mathrm{proj}}=-\sum_{\langle ij\rangle}(\zeta_{ij}\hat f_i^\dagger \hat f_j+\mathrm{H.c.})+
\sum_i(V_i-J_{{\rm ex},i})\hat f_i^\dagger \hat f_i,
\qquad \zeta_{ij}=\langle\chi_i^+|T_{ij}|\chi_j^+\rangle.
\label{eq:Sprojected}
\end{equation}
The corresponding Bloch matrix, with sublattice indices $\alpha,\beta$, is
\begin{align}
[H_{\mathrm{proj}}(\bk)]_{\alpha\alpha}&=V_\alpha-J_{{\rm ex},\alpha},\nonumber\\
[H_{\mathrm{proj}}(\bk)]_{\alpha\beta}&=-2\zeta_{\alpha\beta}\cos(\bk\cdot\mathbf d_{\alpha\beta}),\qquad \alpha<\beta,\nonumber\\
[H_{\mathrm{proj}}(\bk)]_{\beta\alpha}&=[H_{\mathrm{proj}}(\bk)]_{\alpha\beta}^{*}.
\label{eq:SprojectedBloch}
\end{align}
In particular, a single common energy shift $+J$ gives
\begin{equation}
\widetilde H_{\mathrm{proj}}=H_{\mathrm{proj}}+JI_4=H_{\rm hop}+\operatorname{diag}(V_1,V_2,V_3,V_4+\delta),
\qquad \delta=(1-r)J.
\label{eq:Sdetuning}
\end{equation}
The shifted band energies are $\epsilon_n=E_n+J$. At $V_i=0$, $\delta\simeq1.911|t|$ for $J=4|t|$ and $3.822|t|$ for $J=8|t|$.

For $V_i=0$, the leading finite-exchange correction to $H_{\mathrm{proj}}$ scales as $\tau^2/J_{\min}$:
\begin{equation}
H_{\rm low}=H_{\mathrm{proj}}+O(\tau^2/J_{\min}),\qquad
\tau=\max_{ij}(|t_{ij}|+|\lambda_{ij}|),\qquad J_{\min}=\min_iJ_{{\rm ex},i};
\end{equation}
third-order terms scale as $O(\tau^3/J_{\min}^2)$. Figure~\ref{fig:Sbands} compares the gaps of the projected and original models.

At zero SOC, $\zeta_{ij}=t_{ij}\langle\chi_i^+|\chi_j^+\rangle$ and $|\langle\chi_i^+|\chi_j^+\rangle|^2=(1+\bn_i\cdot\bn_j)/2$. The phase of a product of overlaps around an oriented triangle is half of its oriented solid angle.

Writing $u_{ij}=\langle\chi_i^+|\chi_j^+\rangle$, the exchange directions in Eq.~\eqref{eq:Sprofile} give $|u_{12}|=\sqrt{\frac9{11}}\simeq0.905$, $|u_{14}|=\sqrt{\frac{1+5/\sqrt{33}}2}\simeq0.967$.
By threefold symmetry, these are the magnitudes on all kagome--kagome and kagome--triangular bonds, respectively. The projected kagome--triangular hopping magnitude is even slightly larger than the kagome--kagome one. The small triangular-site weight cannot be attributed to a reduced hopping magnitude; it reflects instead quantum interference between different hopping paths. Sections~\ref{sec:Suniform} and \ref{sec:SrealOverlap} study the roles of detuning and complex overlap phases in opening the filling-one gap.

\subsection{Embedded and periodic gauges}
The following Bloch-basis conventions are used in evaluating the Berry
curvature and Hall integers of main-text Eqs.~(4)--(6).
We distinguish the embedded and periodic cell Bloch bases. In the embedded basis, which retains the intracell positions $\mathbf r_\alpha$,
\begin{equation}
H(\bk+\mathbf b_a)=D_aH(\bk)D_a^\dagger,
\qquad (D_a)_{\alpha\alpha}=e^{\ii\mathbf b_a\cdot\mathbf r_\alpha}.
\label{eq:Ssewing}
\end{equation}
Here $a=1,2,3$ labels reciprocal directions; in the original model, $D_a$ acts identically on both spin components. The Hall-response derivation below uses the embedded basis. For the Wilson-loop calculation, we use the periodic cell basis defined by
\begin{equation}
H_{\rm per}(\bk)=W(\bk)H(\bk)W^\dagger(\bk),
\qquad W_{\alpha\alpha}(\bk)=e^{\ii\bk\cdot\mathbf r_\alpha},
\label{eq:Sgauge}
\end{equation}
for which $H_{\rm per}(\bk+\mathbf b_a)=H_{\rm per}(\bk)$. In the original model the transformation is $W\otimes I_2$. The transformation is unitary and therefore leaves the band energies unchanged. Equations~\eqref{eq:Ssewing} and~\eqref{eq:Sgauge} also apply to $H_{\mathrm{proj}}$, with the four-dimensional matrices $D_a$ and $W$.

\section{Bulk spectra and gaps}
\label{sec:Sgaps}
We compare the projected spectrum in Fig.~2 of the main text with the
original Hamiltonian in main-text Eq.~(1), at the same filling and
exchange profile.
For $m$ filled bands, define the direct and indirect gaps separately:
\begin{equation}
\Delta^{\rm dir}_m=\min_{\bq}[E_{m+1}(\bq)-E_m(\bq)],\qquad
\Delta^{\rm ind}_m=\min_{\bq}E_{m+1}(\bq)-\max_{\bq}E_m(\bq).
\end{equation}
An insulating state requires $\Delta_m^{\rm ind}>0$. Table~\ref{tab:Sgaps} gives the lowest global gap results used in the figures. We fill one electron per primitive cell, filling the lowest band.
\begin{table}[ht]
\caption{Filling-one results for $t < 0$, $V_i=\lambda=0$, and $J_{{\rm ex},4}=rJ$. Energies are in units of $|t|$, and $\widetilde\mu=(\max\epsilon_1+\min\epsilon_2)/2$. $\bC$ uses the Hall convention of Sec.~\ref{sec:Stopology}.}
\label{tab:Sgaps}
\begin{ruledtabular}\begin{tabular}{llrrrr}
$J/|t|$ & Model & $\Delta^{\rm dir}_1$ & $\Delta^{\rm ind}_1$ & $\widetilde\mu/|t|$ & $\bC$\\
4 & Projected & 0.3697 & 0.1944 & -1.708 & $(1,1,1)$\\
4 & Original & 0.1889 & 0.05806 & -1.84 & $(1,1,1)$\\
8 & Projected & 0.3818 & 0.1944 & -1.708 & $(1,1,1)$\\
8 & Original & 0.2547 & 0.09881 & -1.794 & $(1,1,1)$
\end{tabular}\end{ruledtabular}
\end{table}

\begin{figure}[!tbp]
\centering
\begin{minipage}{0.455\textwidth}\centering (a) Projected $J=4|t|$\\\includegraphics[width=\textwidth]{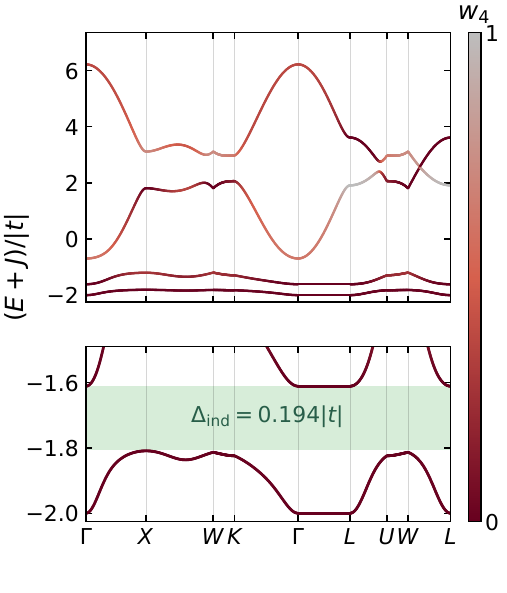}\end{minipage}\hspace{0.035\textwidth}
\begin{minipage}{0.455\textwidth}\centering (b) Original $J=4|t|$\\\includegraphics[width=\textwidth]{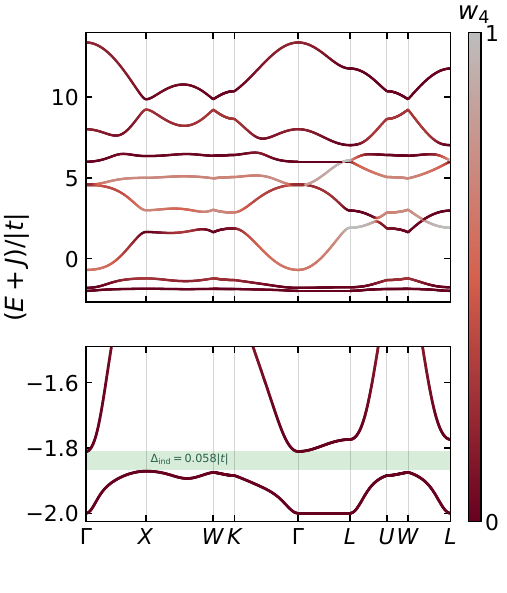}\end{minipage}\\[3pt]
\begin{minipage}{0.455\textwidth}\centering (c) Projected $J=8|t|$\\\includegraphics[width=\textwidth]{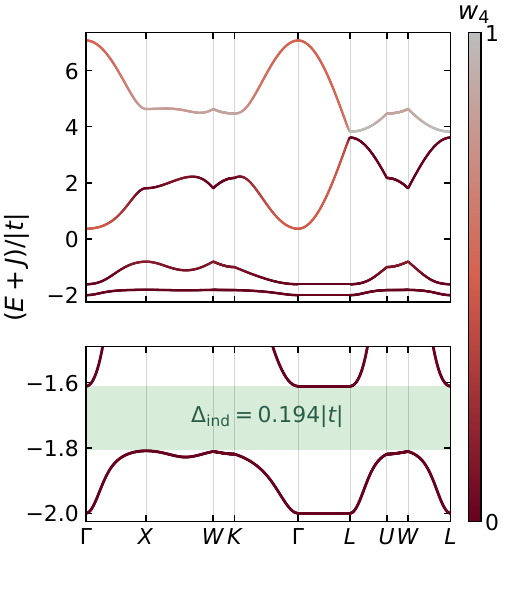}\end{minipage}\hspace{0.035\textwidth}
\begin{minipage}{0.455\textwidth}\centering (d) Original $J=8|t|$\\\includegraphics[width=\textwidth]{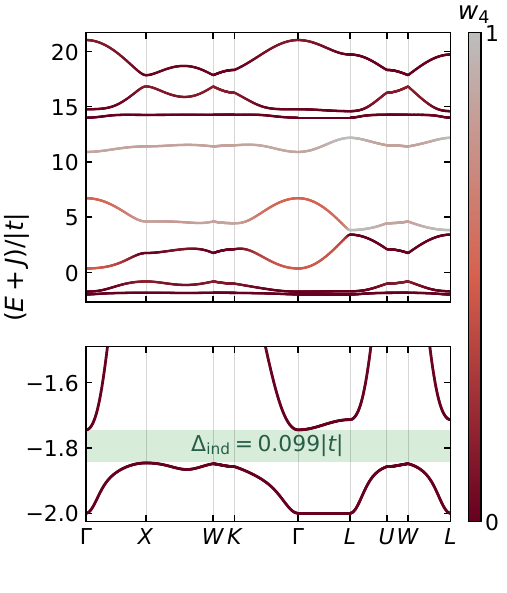}\end{minipage}
\caption{Projected/original comparison of band structure for $J_{{\rm ex},123}=J=4|t|$ (top) and $8|t|$ (bottom), $J_{{\rm ex},4}=rJ$, $t < 0$ and $V_i=\lambda=0$. Panel (a) repeats the main-text reference for comparison. Every panel shows all four or eight bands and a lower enlargement of the filling-one gap; energies are $\epsilon=E+J$. Colors show site-4 probability and green shading marks the global indirect gap. The occupied band at filling one has Chern vector $\mathbf C=(1,1,1)$ in all four cases.}
\label{fig:Sbands}
\end{figure}

The original model preserves the occupied topology but reduces the indirect gap. Figure~\ref{fig:Sbands} shows the full spectra; green shading marks the global indirect gap. The site-4 weight is given by $|\psi_{4n}|^2$ in the projected model and by $\sum_s|\psi_{4sn}|^2$ in the original model.

The original-model extrema were searched numerically from eight separated seeds on $24^3$ and shifted $40^3$ meshes, supplemented by 5000 random momenta per case. Projected checks used shifted $48^3$ and $65^3$ meshes and differential-evolution searches.

\section{Berry curvature, Hall integers, and Wilson loops}
\label{sec:Stopology}
We give the conventions and calculations behind the Berry curvature
formula, Chern triplet, and Hall vector in main-text Eqs.~(4)--(6),
and the response action underlying main-text Eq.~(7).

\subsection{From the three-dimensional Kubo formula to the Berry curvature}
Throughout this subsection, $H=H_{\rm emb}$ is the original Bloch Hamiltonian in the embedded basis and $|u_n\rangle$ are its eigenvectors. Take electron charge $-e$, with $e>0$, $\epsilon_{xyz}=+1$, and $\mathcal A_\alpha=\ii\langle u|\partial_{k_\alpha}u\rangle$. Cartesian indices $\alpha,\beta,\gamma\in\{x,y,z\}$ are summed when repeated. The dc antisymmetric conductivity follows from the Kubo formula~\cite{XiaoChangNiu2010}:
\begin{equation}
\sigma^{{\rm H}}_{\alpha\beta}=e^2\hbar\int_{\rm BZ}\frac{\dd^3k}{(2\pi)^3}
\sum_{n\ne m}\frac{(f_n-f_m)\,\mathrm{Im}[v_\alpha^{nm}v_\beta^{mn}]}{(E_n-E_m)^2},
\label{eq:SKuboVelocity}
\end{equation}
Here $f_n=n_F(E_{n\bk}-\mu)$ and $v_\alpha^{nm}=\langle u_n|\hat v_\alpha|u_m\rangle$. The velocity operator $\hat v_\alpha$ has matrix form $v_\alpha=\hbar^{-1}\partial_{k_\alpha}H$ in the embedded Bloch basis. The interband identity
$\langle u_m|\partial_{k_\alpha}u_n\rangle=\hbar v_\alpha^{mn}/(E_n-E_m)$ gives
\begin{align}
\Omega^{n,(\boldsymbol{k})}_{\alpha\beta}
&=-2\hbar^2\,\mathrm{Im}\sum_{m\ne n}\frac{v_\alpha^{nm}v_\beta^{mn}}{(E_n-E_m)^2}
=-2\,\mathrm{Im}\langle\partial_{k_\alpha}u_n|\partial_{k_\beta}u_n\rangle,\nonumber\\
\sigma^{{\rm H}}_{\alpha\beta}
&=-\frac{e^2}{\hbar}\sum_n\int_{\rm BZ}\frac{\dd^3k}{(2\pi)^3}f_n\Omega^{n,(\boldsymbol{k})}_{\alpha\beta}.
\label{eq:SKuboBerry}
\end{align}
At zero temperature, this reduces to main-text Eq.~(4). For an occupied subspace separated from the unoccupied bands,
let
\[
U(\bq)=\bigl(
|u_1(\bq)\rangle,\ldots,|u_{N_{\rm occ}}(\bq)\rangle
\bigr),
\]
where the columns are orthonormal and $N_{\rm occ}$ is
the number of occupied bands.
Its non-Abelian Berry curvature in reduced coordinates is
\[
\Omega^{(q)}_{bc}
=
\frac{\partial k_\alpha}{\partial q_b}
\frac{\partial k_\beta}{\partial q_c}
\Omega^{(k)}_{\alpha\beta}
=
b_{b,\alpha}b_{c,\beta}\Omega^{(k)}_{\alpha\beta},
\]
where $b_{b,\alpha}$ is the $\alpha$-th Cartesian component of the reciprocal lattice vector $\mathbf b_b$. For cyclic permutations of $(a,b,c)=(1,2,3)$, where 1,2,3 label the three primitive reciprocal lattice directions, we define the first Chern number on the two-torus at fixed $q_a$ as:
\begin{equation}
C_a=-\frac{1}{2\pi}
\int_0^1\dd q_b\int_0^1\dd q_c\,
\Tr\Omega^{(q)}_{bc}.
\label{eq:SC}
\end{equation}

We compute the Chern numbers using the
Fukui--Hatsugai--Suzuki method~\cite{FukuiHatsugaiSuzuki2005}.
For the occupied-band frame $U(\bq)$ and grid step
$\Delta_b=\Delta q_b\hat{\mathbf e}_b$, define the normalized
link
\begin{equation}
\mathcal U_b(\bq)=
\frac{\det[U^\dagger(\bq)U(\bq+\Delta_b)]}
{|\det[U^\dagger(\bq)U(\bq+\Delta_b)]|}.
\end{equation}
For links crossing a Brillouin-zone boundary, the forward
neighbor is replaced by $S_{\mathbf G}U(\bq')$, where
$\bq'$ is the wrapped momentum and $S_{\mathbf G}$ is the
sewing matrix in Eq.~\eqref{eq:Ssewing}.

The phase accumulated around a positively oriented
$(b,c)$ plaquette is
\begin{equation}
F_{bc}(\bq)=
\arg\!\left[
\mathcal U_b(\bq)
\mathcal U_c(\bq+\Delta_b)
\mathcal U_b^{-1}(\bq+\Delta_c)
\mathcal U_c^{-1}(\bq)
\right].
\label{eq:SembeddedFHS}
\end{equation}
The Chern number follows by summing over the slice at
fixed $q_a$:
\begin{equation}
C_a=\frac{1}{2\pi}\sum_{q_b,q_c}F_{bc}(\bq),
\qquad (a,b,c)\ \text{cyclic},
\end{equation}
consistent with the sign convention in Eq.~\eqref{eq:SC}.
The embedded and periodic-cell calculations agree, yielding
$\bC=(1,1,1)$ for both models at $J/|t|=4,8$ and zero SOC.

The quantized three-dimensional Hall response is~\cite{Halperin1987,KohmotoHalperinWu1992}
\begin{equation}
\bG_{\rm H}=\sum_aC_a\mathbf b_a,\qquad
\sigma_{\alpha\beta}=\frac{e^2}{2\pi h}\epsilon_{\alpha\beta\gamma}G_{{\rm H},\gamma}.
\end{equation}
For $\bC=(1,1,1)$, this yields $\bG_{\rm H}=(2\pi/a)(1,1,1)$ and $\sigma_{yz}=\sigma_{zx}=\sigma_{xy}=e^2/(ha)$.

\subsection{Wilson-loop and hybrid-Wannier verification}
\begin{figure}[!tbp]
\centering\includegraphics[width=0.91\textwidth]{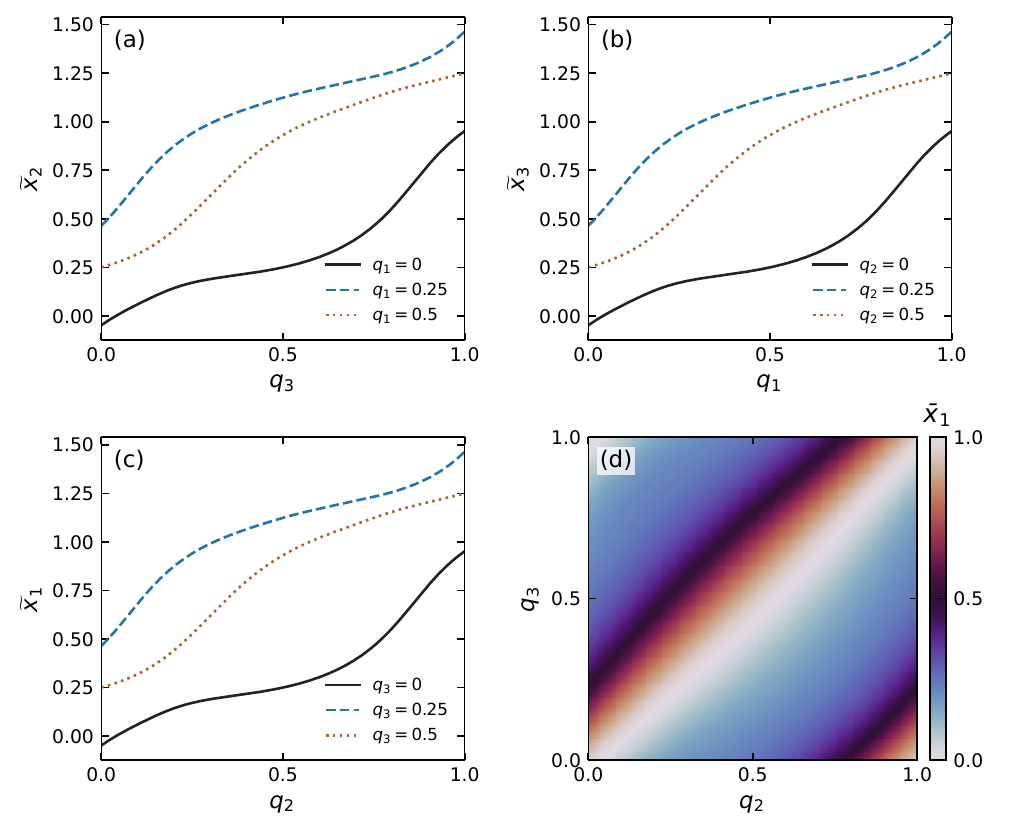}
\caption{Bulk Wilson centers for the projected $J=4|t|$ model with $J_{{\rm ex},4}=4r|t|$, $t < 0$ and $V_i=\lambda=0$. (a--c) Unwrapped centers in the three cyclic constructions; each increases by one. The legends specify the fixed transverse coordinate, with three fixed-$q_3$ cuts in (c). (d) Wrapped $q_1$-loop center on the $(q_2,q_3)$ torus; the cyclic color scale identifies 0 and 1.}
\label{fig:Swilson}
\end{figure}

This subsection gives the Wilson-loop check of the Chern triplet stated after Eq.~(6) of the main text. For the isolated occupied band the Wilson loop is Abelian. Following the hybrid-Wannier formulation~\cite{Coh2009}, define
\begin{align}
\gamma_1(q_2;q_3)&=-\mathrm{Im}\oint_0^1\!\dd q_1\,
\langle u_{\bq}|\partial_{q_1}u_{\bq}\rangle,
&\bar x_1(q_2;q_3)&=\frac{\gamma_1(q_2;q_3)}{2\pi}\pmod1.
\label{eq:SWilsonCenter}
\end{align}
Here $\bar x_1$ is the center in the primitive-cell coordinate $n_1$, conjugate to $q_1$. On a discrete periodic loop,
\begin{equation}
U_{1,j}=\frac{\langle u(q_{1,j},q_2,q_3)|u(q_{1,j+1},q_2,q_3)\rangle}
{|\langle u(q_{1,j},q_2,q_3)|u(q_{1,j+1},q_2,q_3)\rangle|},\qquad
W_1(q_2,q_3)=\prod_jU_{1,j},\qquad
\bar x_1=-\frac{\arg W_1}{2\pi}\pmod1.
\label{eq:SWilsonProduct}
\end{equation}
The calculation uses the periodic cell gauge in Eq.~\eqref{eq:Sgauge}.
\begin{equation}
\Delta_{q_3}\widetilde x_2=C_1,\qquad
\Delta_{q_1}\widetilde x_3=C_2,\qquad
\Delta_{q_2}\widetilde x_1=C_3.
\label{eq:SWilsonTriplet}
\end{equation}
The Wilson-loop spectrum of the projected model, complementing the results in the main text, is shown at Fig.(\ref{fig:Swilson}). We also verified that the projected and original models at $J/|t|=4,8$ give $(C_1,C_2,C_3)=(1,1,1)$ on all sampled slices.

We use 241 points per Wilson loop, 181 points per flow and a $121^2$ torus grid. Independent FHS calculations on $31^2$ and $61^2$ meshes confirm the same Chern number.

\FloatBarrier
\subsection{Effective electromagnetic action}
\label{sec:Saction}
This subsection gives the electromagnetic action mentioned in the main
text and the bulk current entering its surface-continuity equation,
Eq.~(7).
The quantized Hall response is encoded in the crystalline Chern--Simons term~\cite{NissinenVolovik2019}
\begin{equation}
S_{\rm H}[A]=\frac{e^2}{4\pi h}\int\dd^4x\,
\epsilon^{\mu\nu\rho\sigma}G_{{\rm H},\mu}A_\nu\partial_\rho A_\sigma,
\qquad G_{{\rm H},\mu}=(0,\bG_{\rm H}).
\label{eq:SHallAction}
\end{equation}
Here $x^0=t$, $A_\mu=(-\phi,\mathbf A)$, $\epsilon^{\mu\nu\rho\sigma}$ is the totally antisymmetric tensor with $\epsilon^{0123}=-1$, and $F_{\mu\nu}=\partial_\mu A_\nu-\partial_\nu A_\mu$. For variations supported in the bulk, $j^\nu_{\rm H}=\delta S_{\rm H}/\delta A_\nu$ gives
\begin{equation}
j^\nu_{\rm H}=\frac{e^2}{4\pi h}\epsilon^{\mu\nu\rho\sigma}G_{{\rm H},\mu}F_{\rho\sigma},
\qquad \rho_{\rm H}=\frac{e^2}{2\pi h}\bG_{\rm H}\cdot\mathbf B,
\qquad \mathbf j_{\rm H}=\frac{e^2}{2\pi h}\mathbf E\times\bG_{\rm H}.
\label{eq:SHallCurrent}
\end{equation}
The density gives the bulk Streda response. These expressions denote the quantized contributions; nonuniversal polarization and magnetization may also be present. The surface-charge balance in Eq.~(7) of the main text follows by equating the outward normal component of the bulk current to the charge deposited at the surface.

Locally, integration by parts relates Eq.~\eqref{eq:SHallAction} to
\begin{align}
S_\theta[A]&=\frac{e^2}{32\pi^2\hbar}\int\dd^4x\,
\theta\,\epsilon^{\mu\nu\rho\sigma}F_{\mu\nu}F_{\rho\sigma}
=\frac{e^2}{2\pi h}\int\dd t\,\dd^3r\,\theta\,\mathbf E\cdot\mathbf B,\nonumber\\
\theta(\br)&=\theta_0-\bG_{\rm H}\cdot\br\pmod{2\pi},
\qquad \boldsymbol\nabla\theta=-\bG_{\rm H}.
\label{eq:SThetaAction}
\end{align}
The two forms give the same bulk response; on a finite sample, the boundary term from integration by parts must be retained. Since $\bG_{\rm H}$ is reciprocal, a Bravais translation changes $\theta$ by an integer multiple of $2\pi$, preserving lattice translations. Here $\theta$ specifies a static spatial winding, distinct from the constant magnetoelectric angle of a three-dimensional topological insulator~\cite{QiHughesZhang2008,EssinMooreVanderbilt2009,CohVanderbiltMalashevichSouza2011}.

\section{Gap mechanism and parameter dependence}

\subsection{Exchange dependence of the bulk gap}
\label{sec:SexchangeDependence}
\begin{figure}[p]
\centering
\includegraphics[width=0.97\linewidth]{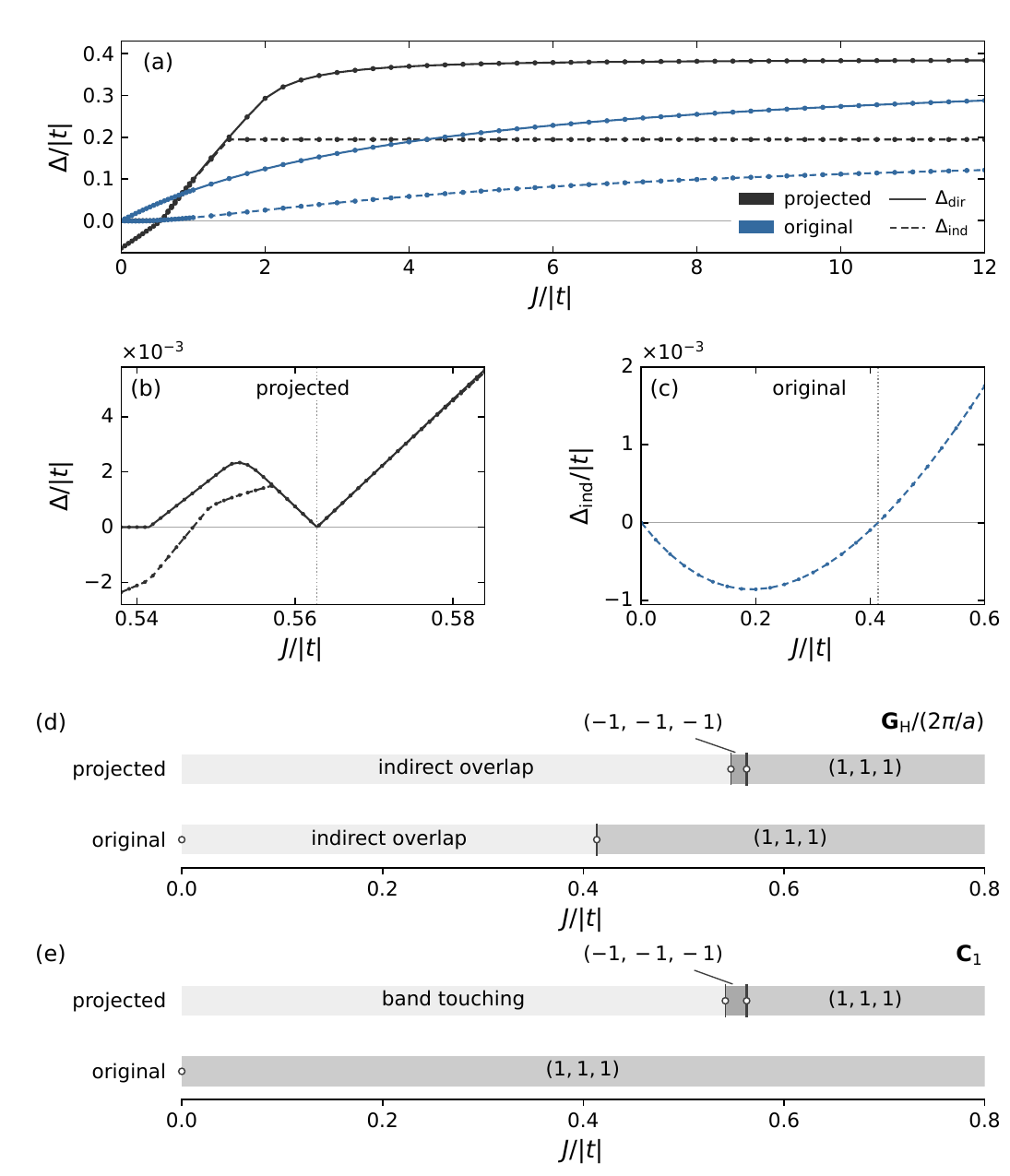}
\caption{\textbf{Exchange dependence of the bulk gaps and topology.}
Equal hopping $t<0$, $J_{{\rm ex},1,2,3}=J$,
$J_{{\rm ex},4}=\sqrt{3/11}J$, and $V_i=\lambda=0$.
(a) Minimum direct (solid) and indirect (dashed) gaps for the projected
(black) and original (blue) models.
(b) Projected small-$J$ enlargement, showing both gaps.
(c) Original-model indirect-gap enlargement.
(d) Indirect-overlap regions and quantized Hall vectors
$\mathbf G_{\rm H}/(2\pi/a)$ in the insulating intervals.
(e) Band-touching regions and the isolated lowest-band Chern vector
$\mathbf C_1$, which remains defined under indirect overlap.
Open circles in (d) and (e) mark zero-gap boundaries. Dots in (a) (b) (c) denote numerical samples connected by straight segments.
In (a), the increment in $J/|t|$ is 0.05 below 1 and 0.25
above 1; in (b) and (c), it is 0.001 and 0.025, respectively.}
\label{fig:SexchangeDependence}
\end{figure}

To see if the physics survives beyond the $J=4|t|$ result in Fig.~2 of the main text and the $J=8|t|$ comparison above, we show in Fig.~\ref{fig:SexchangeDependence} the exchange dependence of
the projected and original Hamiltonians,
Eqs.~\eqref{eq:SprojectedBloch} and~\eqref{eq:SH8}, for
$0\leq J/|t|\leq12$. The parameters are $t<0$,
$J_{{\rm ex},1,2,3}=J$, $J_{{\rm ex},4}=\sqrt{3/11}J$, and
$V_i=\lambda=0$. Here $\Delta_{\rm dir}$ and $\Delta_{\rm ind}$ denote
the lowest band gaps defined above. The lowest-band Chern vector $\mathbf C_1$
remains well defined under indirect overlap as long as $\Delta_{\rm dir}>0$.
In terms of transport, an insulator requires
$\Delta_{\rm ind}>0$, with
$\mathbf G_{\rm H}=\sum_a(\mathbf C_1)_a\mathbf b_a$.

In the original model, the indirect gap opens at
$J/|t|\simeq0.413$, signaling a metallic to insulating transition (the latter being topological with $\mathbf{C}_1=(1,1,1)$. In the projected model, the lowest band has
$\mathbf C_1=(-1,-1,-1)$ between the direct-gap closures at
$J/|t|\simeq0.542$ and $0.563$; its indirect gap opens at
$J/|t|\simeq0.547$. Both gaps close at $J/|t|\simeq0.5628$ and reopen with
$\mathbf C_1=(1,1,1)$.
This weak-$J$ interval lies outside the controlled strong-exchange
projection; the original model has $\mathbf C_1=(1,1,1)$ there.

\subsection{Role of the exchange-induced onsite potential detuning}
\label{sec:Suniform}
The unequal exchange energies in main-text Eq.~(3) produce the
triangular-site detuning $\delta=(1-r)J$, with $r=\sqrt{3/11}$,
after the common energy shift $+J$. To isolate its role in gap
formation, we retain the physical exchange profile
$J_{{\rm ex},1,2,3}=J$, $J_{{\rm ex},4}=rJ$ and the directions
$\bn_i$, and we add a scalar potential $V_4=-\delta$, with
$V_{1,2,3}=0$. This cancels the onsite offset $V_4+\delta$ in
Eq.~\eqref{eq:Sdetuning}, while preserving the texture-induced
complex hoppings. With $t<0$ and $\lambda=0$, the original
control Hamiltonian and its leading local-spin projection are
\begin{align}
\hat H^{\rm comp}
&=-t\sum_{\langle ij\rangle}
(\hat c_i^\dagger\hat c_j+\mathrm{H.c.})
-\sum_iJ_{{\rm ex},i}
\hat c_i^\dagger(\bn_i\cdot\bm\sigma)\hat c_i
-\delta\sum_{i\in4}\hat c_i^\dagger\hat c_i,\nonumber\\
\widetilde{\hat H}_{\mathrm{proj}}^{\rm comp}
&=-t\sum_{\langle ij\rangle}
(u_{ij}\hat f_i^\dagger\hat f_j+\mathrm{H.c.}),
\label{eq:SuniformControl}
\end{align}
where $i\in4$ denotes triangular sites and the tilde denotes
the single-particle energy shift $\epsilon=E+J$.
With this compensating $V_4=-\delta$, the potentials $V_i-J_{{\rm ex},i}=-J$ become equal on every site, and the shifted
projected Hamiltonian contains only the hopping term and is
independent of $J$. Its spectra at $J/|t|=4$ and $8$ therefore
coincide.

\begin{figure}[!tbp]
\centering
\includegraphics[width=0.95\textwidth]
{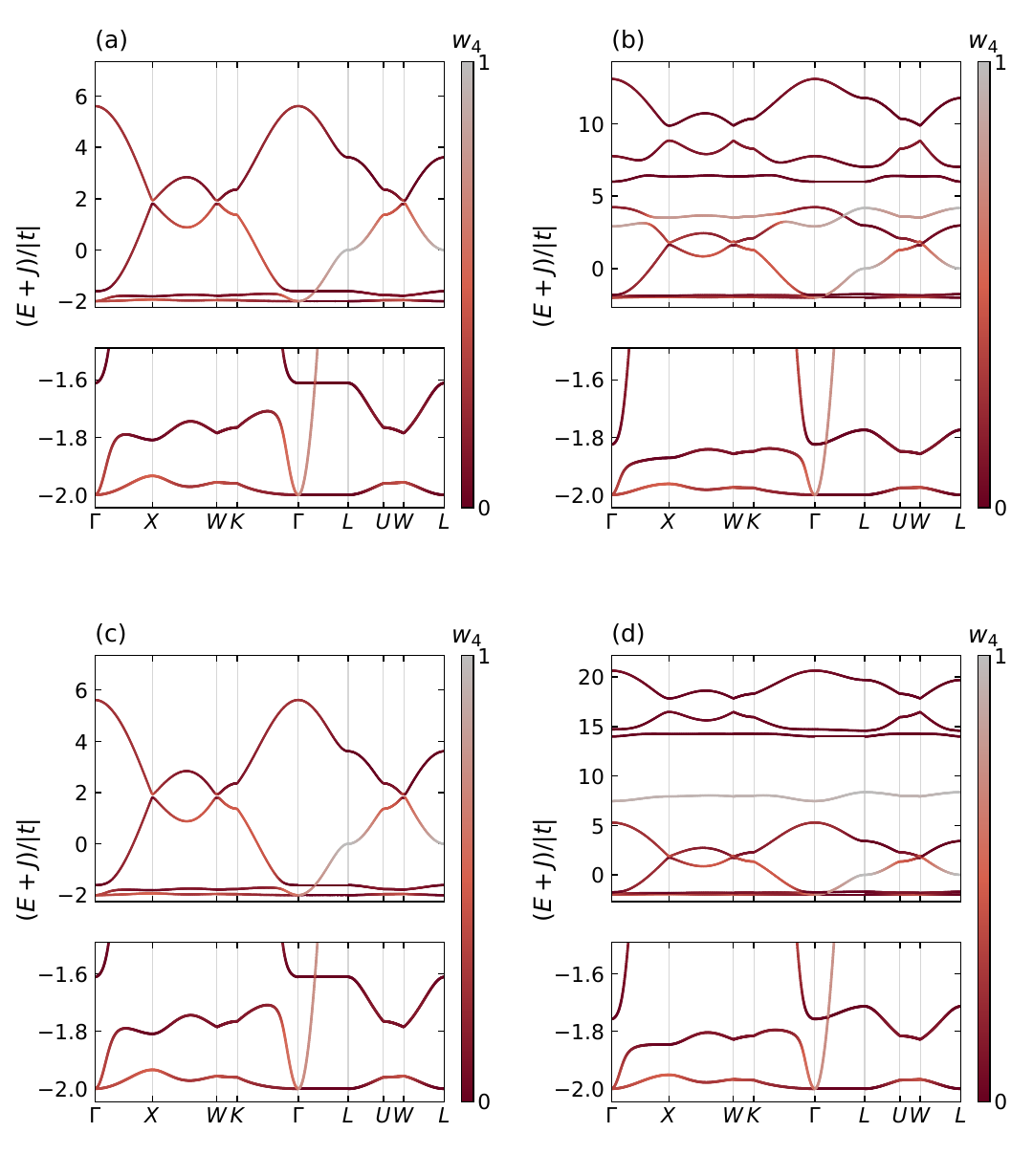}
\caption{Physical exchange profile $J_{{\rm ex},1,2,3}=J$,
$J_{{\rm ex},4}=rJ$, with compensating potential
$V_4=-(1-r)J$, $V_{1,2,3}=0$, $t<0$, and $\lambda=0$.
Projected and original-model spectra are shown for $J=4|t|$
in (a,b) and $J=8|t|$ in (c,d). Each panel shows all bands
and an enlargement of the lowest two, with energies
$\epsilon=E+J$. Colors show site-4 weight, spin summed in
the original model and averaged over exactly degenerate
subspaces. The two projected spectra coincide after the energy shift.}
\label{fig:SuniformBands}
\end{figure}

Fig.~\ref{fig:SuniformBands} shows that lowest-band are in contact at
$\Gamma$, with $\epsilon_1=\epsilon_2=-2|t|$, in both models
at both exchange strengths. We see that the electron occupation of the triangular sites remains low, suggesting that the quasi-layered behavior is primarily driven by hopping phase interference effects. 
Setting $V_4=0$ restores the intrinsic
detuning and the gaps in Table~\ref{tab:Sgaps}, suggesting that the intrinsic onsite potential detuning is an essential contribution to the opening of a global gap between the lowest two bands. 

\subsection{Comparison with a real hopping model}
\label{sec:SrealOverlap}
To test the role of the complex overlap phases in main-text Eq.~(3), we keep the physical unequal-exchange detuning but replace $u_{ij}$ by $|u_{ij}|$ on every bond:
\begin{equation}
\widetilde{\hat H}_4^{\rm real}
=-t\sum_{\langle ij\rangle} \left( |u_{ij}|\hat f_i^\dagger \hat f_j+\mathrm{H.c.} \right)
+\delta\sum_{i\in4}\hat f_i^\dagger \hat f_i,
\qquad \delta=(1-r)J.
\label{eq:SrealOverlapControl}
\end{equation}
The sum $i\in4$ runs over triangular sites. This controlled comparison model preserves each hopping magnitude, the sign $t<0$, and the onsite energies, while removing the texture-induced complex overlap phases. 

\begin{figure}[!tbp]
\centering
\includegraphics[width=0.95\textwidth]{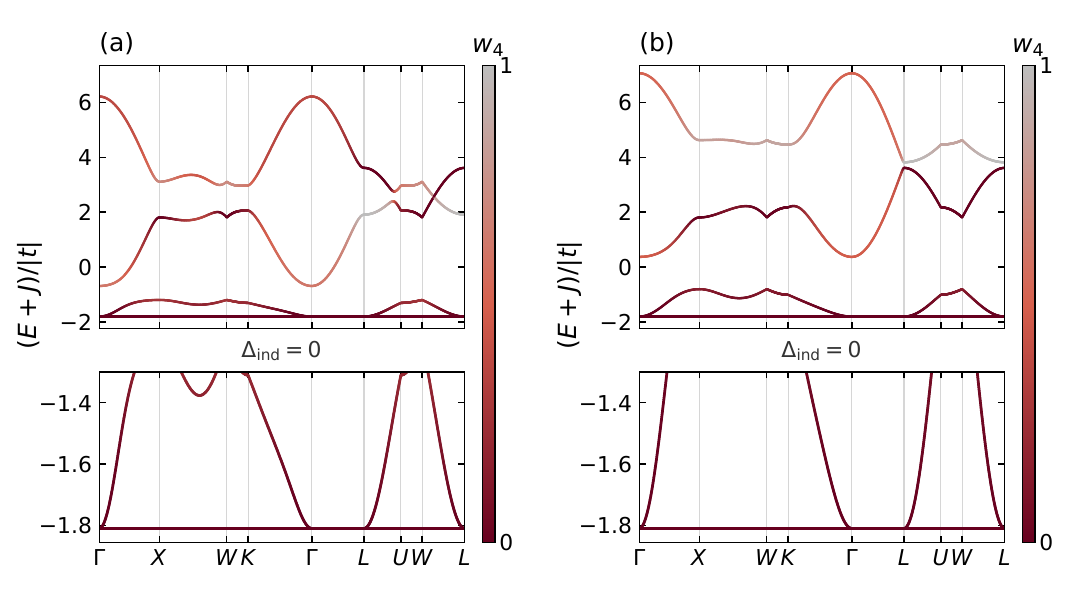}
\caption{Projected real-overlap control, Eq.~\eqref{eq:SrealOverlapControl}, for (a) $J=4|t|$ and (b) $J=8|t|$. The unequal exchange $J_{{\rm ex},123}=J$, $J_{{\rm ex},4}=rJ$ and all hopping magnitudes are retained; only the overlap phases are removed. Both panels show the full four-band spectrum and a lower enlargement, with the same site-$4$ color scale as Fig.~\ref{fig:Sbands}. The flat lowest band touches the second band along $\Gamma$--$L$, so there is no filling-one gap despite the onsite detuning.}
\label{fig:SrealOverlapBands}
\end{figure}

For both $J=4|t|$ and $8|t|$, the lowest band is exactly flat at $\epsilon_1=-2|t|\sqrt{9/11}$ and touches the second band along $\Gamma$--$L$ [Fig.~\ref{fig:SrealOverlapBands}]. Both the direct and indirect filling-one gaps vanish. Away from the contacts, this flat band has zero site-4 weight: amplitudes on kagome sites cancel the hopping into triangular sites. Consequently, detuning and kagome localization alone do not produce the insulating phase. The physical model combines the detuning with the complex overlap phases to open the gap over the lowest-lying quasi-layered band. 
Along \(\Gamma\!-\!L\), two kagome states decouple from site $4$ and remain degenerate at \(\epsilon=-2|t|\sqrt{9/11}\), closing the lowest band gap. The lowest-band contact persists throughout $0\leq t'/t\leq1$
for both $J/|t|=4,8$, including the decoupled-layer limit
$t'=0$.
With the complex overlaps retained, the projected model
remains gapped throughout the same interval.

\subsection{Connection to decoupled Chern layers}
We give an explicit interpolation supporting the main-text statement
that the insulating state is in the same topological class as stacked
QH layers. 
\begin{figure}[htbp]
\centering
\includegraphics[width=0.95\textwidth]{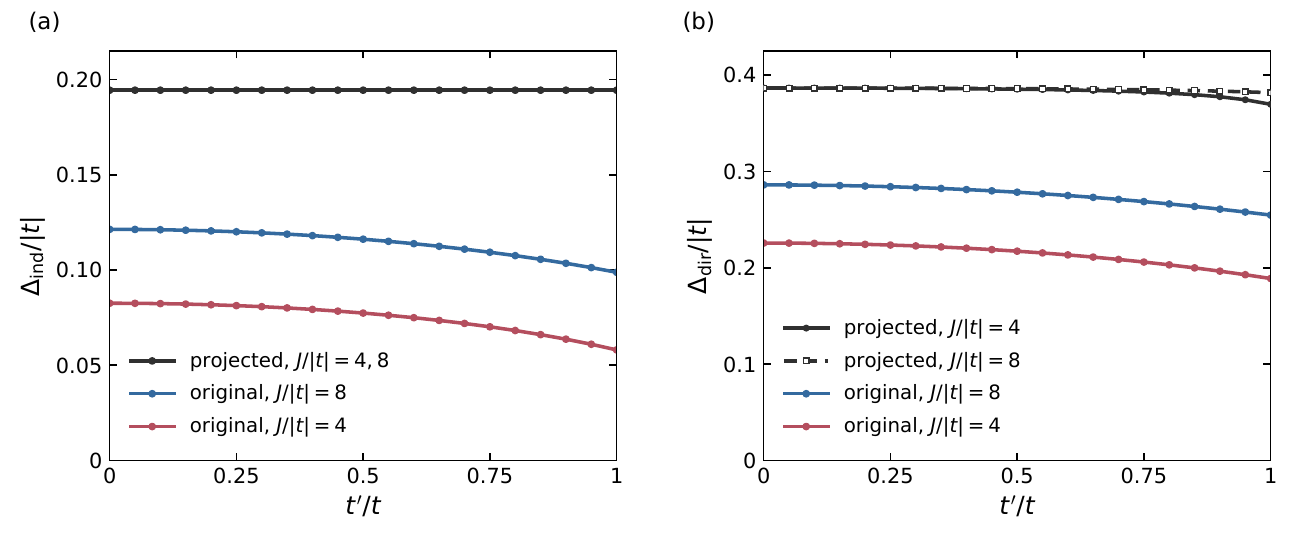}
\caption{(a) Global indirect gap and (b) minimum direct gap over the
three-dimensional Brillouin zone along the interpolation
$s=t'/t$, with the physical exchange fields fixed
and $V_i=\lambda=0$.
The endpoints $s=0$ and $1$ correspond to decoupled
kagome layers and the isotropic pyrochlore model,
respectively. The projected indirect gaps for
$J/|t|=4,8$ coincide; their direct gaps are shown separately.
Markers indicate numerically refined full-zone extrema.}
\label{fig:Slayerconnection}
\end{figure}
We only vary the kagome--triangular hopping $t'=st$,
with $0\le s\le1$, keeping the intrakagome hopping
$t<0$, effective exchange directions, and exchange
magnitudes $J_{{\rm ex},1,2,3}=J$ and
$J_{{\rm ex},4}=rJ$ fixed. At $V_i=\lambda=0$,
\begin{equation}
\hat H(s)=\hat H_{\rm kag}+\hat H_{\rm tri}
+s\hat H_{\rm int},
\end{equation}
where $\hat H_{\rm int}$ contains the
kagome--triangular hopping of the original Hamiltonian. The corresponding $\hat H_{\mathrm{proj}}(s)$ is defined by the same rescaling of its kagome--triangular hoppings.
At $s=0$, the kagome planes are decoupled and the
triangular sites are isolated; $s=1$ restores equal
bare hopping on every bond. For both the projected
and original models at $J/|t|=4,8$, the global indirect
gap remains positive throughout this interpolation
[Fig.~\ref{fig:Slayerconnection}], preserving
$\bC=(1,1,1)$ of the filled lowest band. In the projected model, the indirect gap remains unchanged and is the same for $J/|t|=4,8$, since the lowest unoccupied state and the highest occupied state both have zero site-$4$ weight. The minimum direct gap varies along the interpolation and is shown separately in Fig.~\ref{fig:Slayerconnection}(b).

\section{Slab construction and supplementary surface spectra}
\label{sec:Ssurface}
We specify the slab geometry and the surface-weight and spectral-intensity
definitions used in Figs.~3 and~4 of the main text, and compare the
projected and original-model spectra. Following the bulk--boundary relation
in main-text Eq.~(7), the net surface chirality is directed along $\bG_{\rm H}\times\hat{\mathbf m}$. For the face with outward normal $\hat{\mathbf m}=(-1,1,1)/\sqrt3$, this gives the in-plane direction $[0\bar1 1]$; the opposite face has the opposite chirality.

The slab encompasses $N=28$ complete primitive cells along the $\mathbf a_1$ direction, and periodicity in $\mathbf a_2, \mathbf a_3$. The prescribed bulk exchange fields are retained at the boundary. The faces are parallel to the kagome $234$ planes of the pyrochlore lattice. The top face of the slab (red) has outward normal $(1,-1,-1)/\sqrt3$, and the opposite face (blue) has normal $(-1,1,1)/\sqrt3$. There are $4N=112$ projected or $8N=224$ original-model levels at each $(q_2,q_3)$. 
The fixed-$q_3$ cuts sweep $q_2$ at $q_3=-1/4,0,1/4$.

\begin{figure}[!tbp]
\centering
(a) Original $J=4|t|$\\\includegraphics[width=0.90\textwidth]{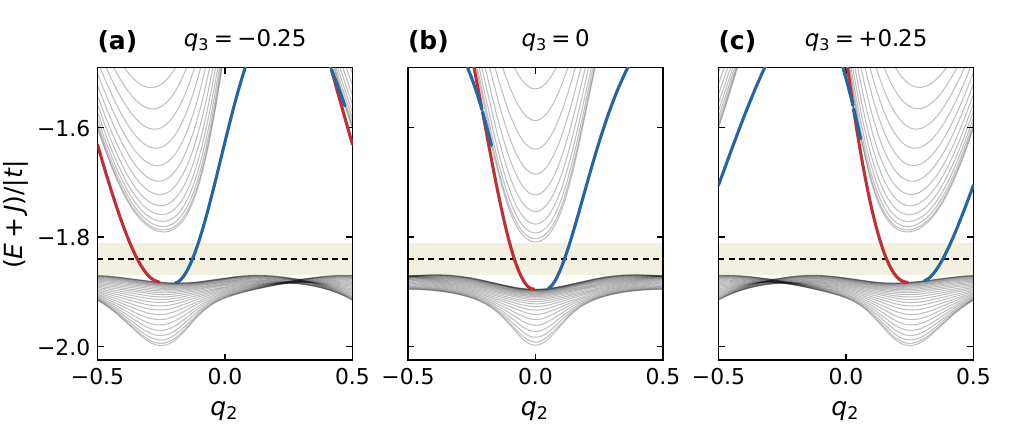}\\[3pt]
(b) Projected $J=8|t|$\\\includegraphics[width=0.90\textwidth]{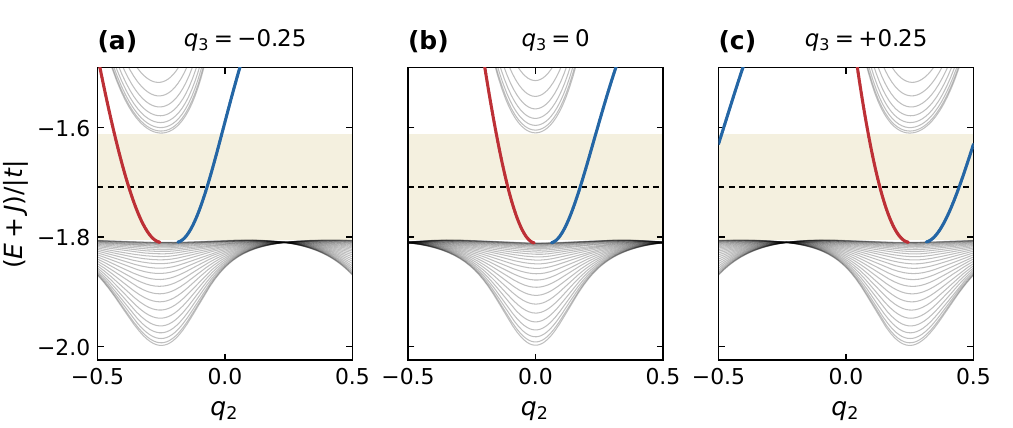}\\[3pt]
(c) Original $J=8|t|$\\\includegraphics[width=0.90\textwidth]{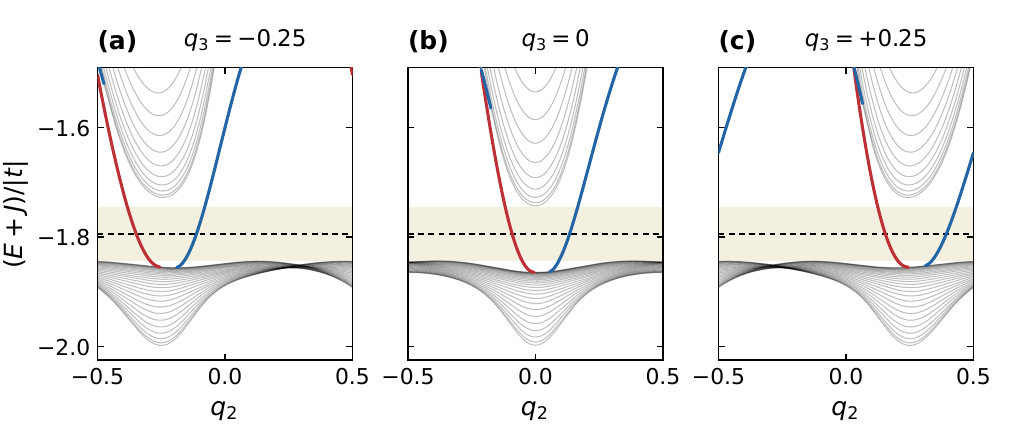}
\caption{Side-surface spectral-flow comparisons at $V_i=\lambda=0$, $t<0$, and $J_{{\rm ex},4}=rJ$. Each row contains $q_3=-1/4,0,1/4$ from left to right, with $q_2$ on the horizontal axis. The red top-face and blue bottom-face crossings have opposite slopes. Energies include the common $+J$ shift.}
\label{fig:Ssurfcuts}
\end{figure}

Let $\hat P_f^{(3)}$ project onto the outermost three unit-cell layer of face $f$ (top/bottom). The finite-slab energy spectrum plots color a state on that face when $w_{fn}=\langle\psi_n|\hat P_f^{(3)}|\psi_n\rangle> 1/2$. Gray curves are all finite-slab energy band. Figure~\ref{fig:Ssurfcuts} supplements the projected $J=4$ cuts in the main text with the original $J=4$ and both $J=8$ cases. Yellow shading marks the bulk indirect gap for each model; the chemical potentials are in Table~\ref{tab:Sgaps}.

The Fermi-surface plots sum over all slab states, and plot the Lorentzian-broadened surface spectral weight derived
from the retarded Green's function:
\begin{equation}
A_f(\bq;\widetilde\mu)
=-\eta\,\mathrm{Im}\,\mathrm{Tr}
\left[P_fG^R(\bq,\widetilde\mu)\right]
=\sum_{n=1}^{dN}w_{fn}(\bq)
\frac{\eta^2}
{[\epsilon_n(\bq)-\widetilde\mu]^2+\eta^2},
\end{equation}
where $G^R(\bq,\omega)
=[\omega-H_{\mathrm{slab}}(\bq)+i\eta]^{-1}$,
$P_f$ projects onto surface $f$, and $\eta=0.01|t|$, $d=4$ or $8$.
We normalize the spectral weight as
$\widetilde A_f=A_f/M_J$, $M_J$ is the maximum of $A_f$ over both faces and both models at fixed $J$. $M_4\simeq1.084$ and $M_8\simeq1.041$ for the $J=4|t|$ and $J=8|t|$ respectively.

The contours $\epsilon_{28}(q_2,q_3)=\widetilde\mu$ are
obtained by solving for the crossings at fixed $q_3$ and
joining the resulting points by linear interpolation.
They bound a region with 28 occupied slab levels;
the complementary region has 27.
Each face supports one noncontractible loop, winding
as $(1,1)$ when oriented along increasing $q_3$.
The two faces have opposite electron chirality.

\section{Robustness toward weak SOC}
\label{sec:Ssoc}
In the main text we mention that the results presented in our work are qualitatively stable against SOC. 
Here present a detailed study of this robustness by adding the bond SOC defined above
to the original Hamiltonian of main-text Eq.~(1).

Table~\ref{tab:Ssoc} lists the filling-one indirect gaps for a uniform SOC strength $\lambda$ in Eq.~\eqref{eq:SH8}. The sign of $\lambda$ is physical once the oriented axes and hopping sign have been fixed.
\begin{table}[ht]
\caption{Original-model indirect gaps $\Delta^{\rm ind}_1/|t|$ for $t<0$, $V_i=0$, and $J_{{\rm ex},4}=rJ$. A negative indirect gap denotes band overlap.}
\label{tab:Ssoc}
\begin{ruledtabular}\begin{tabular}{crrrr}
$J/|t|$ & $\lambda/|t|=-0.02$ & $\lambda/|t|=0$ & $\lambda/|t|=+0.01$ & $\lambda/|t|=+0.02$\\
4 & 0.1268 & 0.05806 & 0.02268 & -0.01367\\
8 & 0.1664 & 0.09881 & 0.06443 & 0.02954
\end{tabular}\end{ruledtabular}
\end{table}
The filling-one Chern-insulating phase is therefore stable against weak SOC: for $J/|t|=4,8$, we find $\Delta_1^{\rm ind}>0$ and $\bC=(1,1,1)$ throughout $|\lambda|\leq0.01|t|$. At $J/|t|=4$, $\lambda=0.02|t|$ produces indirect band overlap.

\section{Stability against nonlocal Kondo exchange}
\label{sec:Snonlocal}
We consider here the effects of the nonlocal Kondo exchange $J_K'$ in main-text Eq.~(1),
\begin{equation}
\delta\hat H_K=-J_K'\sum_{\ell\in A}
\sum_{\substack{i,j\in\mathcal N_B(\ell)\\i<j}}
\left[c_i^\dagger(\mathbf S_\ell^A\cdot\bm\sigma)c_j+\mathrm{H.c.}\right]
\, ,
\label{eq:SnonlocalRealSpace}
\end{equation}
where $\mathcal N_B(\ell)$ is the set formed by the six B-sites nearest neighbors of
A-site $\ell$, and $i<j$ restricts the sum to the $15$ unordered pairs $i,j \in \mathcal N_B(\ell)$. 
In the Bloch basis of Eq.~\eqref{eq:SH8}, the original eight-band matrix is
\begin{equation}
H(\bk;J_K')=H(\bk;0)+\delta H_K(\bk),
\label{eq:SnonlocalBloch}
\end{equation}
where $H(\bk;0)$ is given by Eq.~\eqref{eq:SH8}.
Writing $x=ak_x/4$, $y=ak_y/4$, and $z=ak_z/4$, and each \(\delta H_{ij}\) as a \(2\times2\) spin matrix. The diagonal blocks of $\delta H_K$ are
\begin{align}
\delta H_{11}&=-2J_K'\bigl[\cos 2(x+y)\mathbf S_2^A+
\cos 2(x+z)\mathbf S_3^A+\cos 2(y-z)\mathbf S_4^A\bigr]\cdot\bm\sigma,
\nonumber\\
\delta H_{22}&=-2J_K'\bigl[\cos 2(x+y)\mathbf S_1^A+
\cos 2(y+z)\mathbf S_3^A+\cos 2(x-z)\mathbf S_4^A\bigr]\cdot\bm\sigma,
\nonumber\\
\delta H_{33}&=-2J_K'\bigl[\cos 2(x+z)\mathbf S_1^A+
\cos 2(y+z)\mathbf S_2^A+\cos 2(x-y)\mathbf S_4^A\bigr]\cdot\bm\sigma,
\nonumber\\
\delta H_{44}&=-2J_K'\bigl[\cos 2(y-z)\mathbf S_1^A+
\cos 2(x-z)\mathbf S_2^A+\cos 2(x-y)\mathbf S_3^A\bigr]\cdot\bm\sigma.
\label{eq:SnonlocalDiagonal}
\end{align}
The upper off-diagonal blocks are
\begin{align}
\delta H_{12}={}&-2J_K'\cos(x-y)(\mathbf S_3^A+\mathbf S_4^A)\cdot\bm\sigma
\nonumber\\
&-2J_K'\bigl[\cos(x+y+2z)\mathbf S_3^A+
\cos(x+y-2z)\mathbf S_4^A\bigr]\cdot\bm\sigma,\nonumber\\
\delta H_{13}={}&-2J_K'\cos(x-z)(\mathbf S_2^A+\mathbf S_4^A)\cdot\bm\sigma
\nonumber\\
&-2J_K'\bigl[\cos(x+2y+z)\mathbf S_2^A+
\cos(x-2y+z)\mathbf S_4^A\bigr]\cdot\bm\sigma,\nonumber\\
\delta H_{14}={}&-2J_K'\cos(y+z)(\mathbf S_2^A+\mathbf S_3^A)\cdot\bm\sigma
\nonumber\\
&-2J_K'\bigl[\cos(2x+y-z)\mathbf S_2^A+
\cos(2x-y+z)\mathbf S_3^A\bigr]\cdot\bm\sigma,\nonumber\\
\delta H_{23}={}&-2J_K'\cos(y-z)(\mathbf S_1^A+\mathbf S_4^A)\cdot\bm\sigma
\nonumber\\
&-2J_K'\bigl[\cos(2x+y+z)\mathbf S_1^A+
\cos(2x-y-z)\mathbf S_4^A\bigr]\cdot\bm\sigma,\nonumber\\
\delta H_{24}={}&-2J_K'\cos(x+z)(\mathbf S_1^A+\mathbf S_3^A)\cdot\bm\sigma
\nonumber\\
&-2J_K'\bigl[\cos(x+2y-z)\mathbf S_1^A+
\cos(x-2y-z)\mathbf S_3^A\bigr]\cdot\bm\sigma,\nonumber\\
\delta H_{34}={}&-2J_K'\cos(x+y)(\mathbf S_1^A+\mathbf S_2^A)\cdot\bm\sigma
\nonumber\\
&-2J_K'\bigl[\cos(x-y+2z)\mathbf S_1^A+
\cos(x-y-2z)\mathbf S_2^A\bigr]\cdot\bm\sigma.
\label{eq:SnonlocalOffDiagonal}
\end{align}
The lower blocks satisfy $\delta H_{ji}=\delta H_{ij}^\dagger$.
The diagonal blocks describe same-sublattice intercell hopping.
The first and second lines of each off-diagonal block describe transfers
between nearest B--B neighbors and the additional shared-A transfers
beyond nearest B--B neighbors, respectively.

\begin{figure}[!tbp]
\centering
\includegraphics[width=0.95\textwidth]{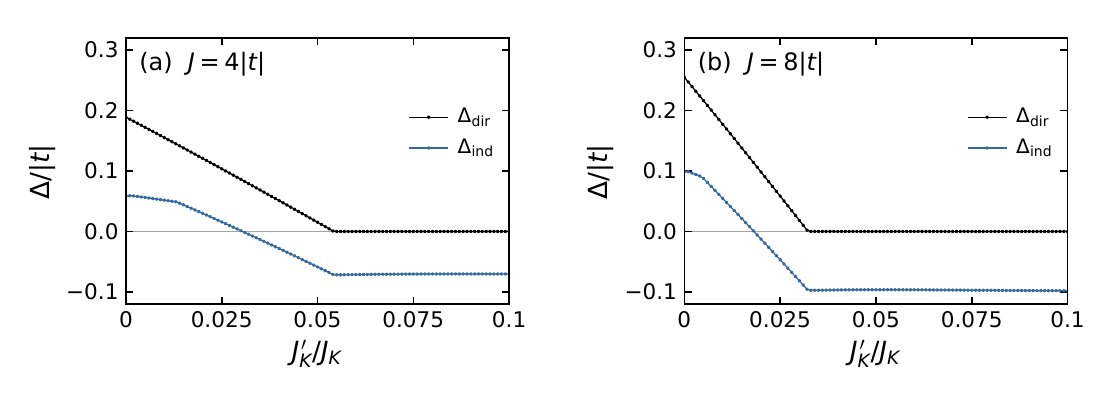}
\caption{Stability against nonlocal Kondo exchange in the original
eight-band model at (a) $J=4|t|$ and (b) $J=8|t|$, with
$t<0$, $J_{{\rm ex},4}=\sqrt{3/11}J$, and $V_i=\lambda=0$.
Black and blue curves show the minimum direct and indirect gaps,
respectively; a negative indirect gap denotes band overlap.
Dots are actual data calculated at spacing $\Delta(J_K'/J_K)=0.001$,
joined by straight lines. 
Band extrema are refined over the full Brillouin zone from $21^3$
meshes, with independent shifted-$31^3$ checks at selected ratios
and near the indirect-gap closures.}
\label{fig:SnonlocalGaps}
\end{figure}

Fig.~\ref{fig:SnonlocalGaps} shows the lowest-band direct and indirect
gaps at $J/|t|=4,8$, with $t < 0$ and $V_i=\lambda=0$.
The magnetic texture and onsite magnitudes
$J_{{\rm ex},1,2,3}=J=2J_KS\sqrt{11/3}$ and
$J_{{\rm ex},4}=\sqrt{3/11}J$ are held fixed while $J_K'/J_K$ varies.
The initial insulating intervals end at
$J_K'/J_K\simeq0.03044$ and $0.01814$, respectively.
At these indirect-gap closures, the direct gaps are
$0.08447|t|$ and $0.11300|t|$.
The lowest-band Chern triplet remains $\bC=(1,1,1)$
until the direct gap closing is reached.

These results show that the 3D QH regime discussed in the main text is robust to weak non-local Kondo exchange. On general grounds we have no reason to expect $J_K' \ll J_K$. The angular dependence of orbital overlap can significantly affect the strength of exchange terms, and materials where $J_K' \ll J_K$ might be possible. It will be interesting to see what other phases and properties these systems might exhibit in the regime $J_K' \sim J_K$; we leave this question to future investigations.

\section{Hartree--Fock calculations}
\label{sec:Shf}
We perform Hartree--Fock calculations for the $J=4|t|$, $J_{{\rm ex},4}=4r|t|$, equal-hopping, zero-SOC Hamiltonian, with $V_i=0$, one electron per cell and $T=0$. The calculation allows mixing among all four projected bands while preserving primitive-cell translations.

In this section, $\alpha,\beta,\gamma=1,\ldots,4$ label the B-site sublattices, with one local orbital per sublattice. The intersite interaction is given by
\begin{equation}
\hat H_{\rm int}=\frac12\sum_{(\mathbf R,\alpha)\ne(\mathbf R',\beta)}
V(|\mathbf R+\mathbf r_\alpha-\mathbf R'-\mathbf r_\beta|)
(\hat n_{\mathbf R\alpha}-\bar n_\alpha)(\hat n_{\mathbf R'\beta}-\bar n_\beta),
\qquad \bar n_\alpha=\tfrac14.
\end{equation}
The potential is $V(r)=g/r$ in numerical units $a=|t|=1$. Unscreened Coulomb uses neutral Ewald summation, removing the divergent total $\mathbf Q+\mathbf G=0$ mode, retaining the finite
$\mathbf Q=0$ terms.

Let $\hat f_{\mathbf R\alpha}$ annihilate an electron in the
locally aligned spin state on sublattice $\alpha$.
We use the periodic cell basis,
$\hat f_{\bk\alpha}=N_c^{-1/2}\sum_{\mathbf R}
e^{-\ii\bk\cdot\mathbf R}\hat f_{\mathbf R\alpha}$,
where $N_c$ is the number of primitive cells, and define
\begin{equation}
P_{\alpha\beta}(\bk)
=\langle\hat f^\dagger_{\bk\beta}\hat f_{\bk\alpha}\rangle,
\qquad
n_\alpha=\frac{1}{N_k}\sum_{\bk}P_{\alpha\alpha}(\bk).
\end{equation}
The self-consistent HF Hamiltonian is
\begin{align}
F_{\alpha\beta}(\bk)
={}&h_{\alpha\beta}(\bk)
+\delta_{\alpha\beta}\sum_\gamma
V_{\alpha\gamma}(0)(n_\gamma-\bar n_\gamma)
\nonumber\\
&-\frac{1}{N_k}\sum_{\mathbf Q}
V_{\alpha\beta}(\mathbf Q)
P_{\alpha\beta}(\bk-\mathbf Q),
\label{eq:SHFmatrix}
\end{align}
where $h(\bk)$ is the projected noninteracting Hamiltonian
in the same basis, $N_k$ is the number of sampled momenta,
$\bar n_\alpha=1/4$, and
\begin{equation}
V_{\alpha\beta}(\mathbf Q)
=\sum_{\mathbf R}'e^{+\ii\mathbf Q\cdot\mathbf R}
V(|\mathbf R+\mathbf r_\beta-\mathbf r_\alpha|).
\label{eq:SHFvertex}
\end{equation}
The prime excludes the identical site.
We iterate $F$ and $P$ to self-consistency.

\begin{figure}[!tbp]
\centering\includegraphics[width=0.95\textwidth]{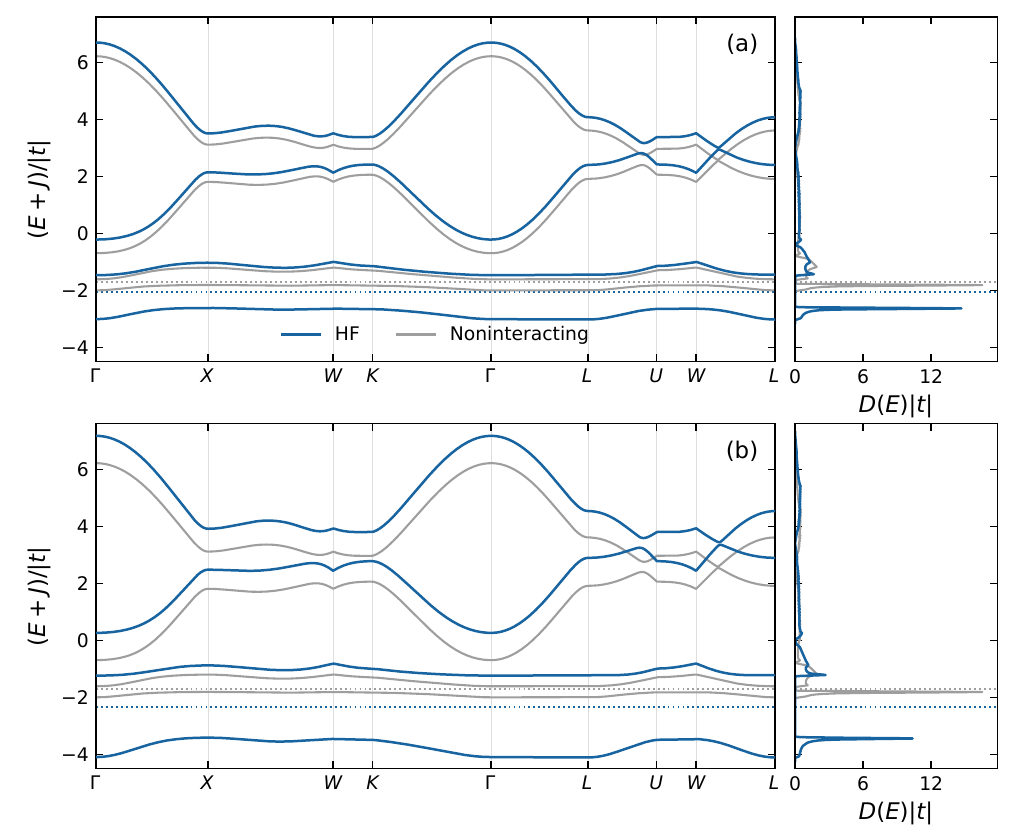}
\caption{Projected four-band spectra and DOS at filling one, for unscreened Coulomb coupling (a) $g=0.5$ and (b) $g=1$. Blue: self-consistent HF; gray: noninteracting, with the same $J=4|t|$, $J_{{\rm ex},4}=4r|t|$, with equal hopping $t=-|t|$ and $V_i=\lambda=0$. Dotted lines show each model's gap-centered chemical potential. Both spectra use the energy shift $E+J$.}
\label{fig:SHFbandsDOS}
\end{figure}

Figure~\ref{fig:SHFbandsDOS} compares the projected $g=0.5,1$ spectra with $g=0$ at the same $24^3$ self-consistent mesh. All four projected bands can mix, while only the lowest self-consistent band is occupied. The corresponding indirect gaps are $1.149|t|$ and $2.174|t|$, compared with $0.1944|t|$ without interactions. The DOS is
\begin{equation}
D(E)=\frac1{N_k}\sum_{n,\bk}\delta_\eta(E-\epsilon_{n\bk}),\qquad
\int D(E)\,\dd E=4,
\end{equation}
where $\delta_\eta$ is a normalized Gaussian $
\delta_\eta(x)=
\frac{1}{\sqrt{2\pi}\eta}
\exp\left(-\frac{x^2}{2\eta^2}\right),
$ of standard deviation $\eta=0.015|t|$. The DOS is evaluated on a $48^3$ mesh using the converged $24^3$ Fock Hamiltonian. Gaussian tails inside the gap are broadening artifacts.

For comparison, Table~\ref{tab:Shf} also gives results for
the original Hamiltonian, treating all eight bands
self-consistently with the same density--density interaction
and filling. The site densities are spin summed, with
the same background $\bar n_\alpha=1/4$.

\begin{table}[ht]
\caption{HF indirect gaps in units of $|t|$ for unscreened
Coulomb interactions, from self-consistent solutions on
$40^3$ meshes. The occupied band has $\bC=(1,1,1)$
in both models for all parameters shown.}
\label{tab:Shf}
\begin{ruledtabular}
\begin{tabular}{rrr}
$g$ & Projected & Original\\
0.5 & 1.1873 & 1.0129\\
1   & 2.251  & 2.075
\end{tabular}
\end{ruledtabular}
\end{table}
The bare gaps are $0.1944|t|$ and $0.05806|t|$ in the projected and original models. HF increases the gap for all parameters studied, while preserving $\bC=(1,1,1)$. For the original model at $g=1$, the indirect gap is $1.999,2.046,2.075|t|$ on $24^3,32^3,40^3$ meshes,
respectively.

Self-consistency was solved on $12^3,16^3,24^3$ meshes
using four initial states at each interacting coupling.
Larger-mesh calculations were initialized from
smaller-mesh solutions: $40^3$ at $g=0.5$, and
$32^3,40^3$ at $g=1$.
The Chern numbers were evaluated on $24^3$ and $40^3$
meshes of the converged Fock Hamiltonian.
Band extrema were refined from multiple starting points.

These Hartree--Fock calculations show that Coulomb interactions enhance the indirect gap at $J_K'=0$ and preserve $\bC_1=(1,1,1)$ for the parameters studied.
Whether interactions extend the stability range against nonlocal exchange, and which phases occur for
$J_K' \sim J_K$, remain open questions for future investigations.

\FloatBarrier

\end{document}